\documentclass{emulateapj}

\newcommand{\const}{{\rm const.}}
\newcommand{\cmcms}{\ {\rm cm^2}~{\rm s^{-1}}}
\newcommand{\eV}{\ {\rm eV}}
\newcommand{\cm}{\ {\rm cm}}
\newcommand{\pc}{\ {\rm pc}}

\def\highlight#1{{#1}}

\usepackage{amsmath}
\usepackage{natbib}
\usepackage{soul}

\usepackage{color}
\usepackage{url}

\shorttitle{Model of PWNe with the diffusion back-reaction}
\shortauthors{ISHIZAKI \textit{ET AL}.}

\begin{document}


\title{
	\highlight{Outflow and Emission Model of Pulsar Wind Nebulae with the Back-reaction of Particle Diffusion}
}


\author{Wataru Ishizaki\altaffilmark{1}, Katsuaki Asano\altaffilmark{1} and Kyohei Kawaguchi\altaffilmark{1}}
%
%


\altaffiltext{1}{Institute for Cosmic Ray Research, The University of Tokyo, 5-1-5 Kashiwa-no-ha, Kashiwa City, Chiba, 277-8582, Japan}


\begin{abstract}
	We present a new pulsar wind nebula (PWN) model solving both advection and diffusion of non-thermal particles in a self-consistent way to satisfy the momentum and energy conservation laws.
	Assuming spherically symmetric (1--D) steady outflow, we calculate the emission spectrum integrating over the entire nebula and the radial profile of the surface brightness.
	We find that the back reaction of the particle diffusion modifies the flow profile.
	The photon spectrum and the surface brightness profile are different from the model calculations without the back reaction of the particle diffusion.
	Our model is applied to the two well-studied PWNe 3C 58 and G21.5-0.9.
	By fitting the spectra of these PWNe, we determine the parameter sets and calculate the radial profiles of X-ray surface brightness.
	For both the objects, obtained profiles of X-ray surface brightness and photon index are well consistent with observations.
	\highlight{
	Our model suggests that particles escaped from the nebula significantly contribute to the gamma-ray flux.
	}
	A $\gamma$-ray halo larger than the radio nebula is predicted in our model.
\end{abstract}


\section{INTRODUCTION}\label{sec:intro}

Pulsar wind nebulae (PWNe) are extended objects around a rotation-powered pulsar with a size of about a few pc,
and their emission spectrum extends from radio to $\gamma$-ray \citep[See review for][]{2006ARA&A..44...17G}.
The emission is due to synchrotron radiation and inverse Compton scattering (ICS) by electrons and positrons accelerated at the termination shock generated by the interaction between the supernova remnant (SNR) and the pulsar wind
\citep{1974MNRAS.167....1R,1992ApJ...396..161D}.
Based on this idea, \citet{1984ApJ...283..694K,1984ApJ...283..710K} established a 1D-steady magnetohydrodynamical (MHD) model (hereafter KC model) of PWNe.

Although the KC model has been accepted as a standard model of PWNe,
some problems in the KC model have been raised by morphology research with high angular resolution observations in X-ray.
The X-ray photon spectrum becomes gradually softer as the radius increases \citep{2004ApJ...616..403S,2005AdSpR..35.1099M}.
\citet{Rey03} pointed out that the gradual softening is incompatible with the KC model, which predicts sudden softening at a certain radius.
\citet{2004ApJ...616..403S} also presented the radial profile of the photon index in 3C 58,
and showed that the predicted profile in the KC model deviates from the observation.
\citet{ishizaki17} applied the KC model to G21.5-0.9 and 3C 58, and showed that the KC model has severe difficulty to reproduce both the entire spectrum and the surface brightness profile simultaneously.

Since the problems in the KC model have been claimed,
several authors have discussed improvement of the KC model.
In the KC model, particles are simply advected with the spherical wind.
However, \citet{2014MNRAS.438..278P} performed a three-dimensional full MHD simulation and showed that PWNe are in very disturbed state.
\citet{2003MNRAS.346..841S} suggested the existence of turbulent magnetic field in the nebula by using a semi-analytic method.
In such disturbed plasma, the particle diffusion can be an important process.
\citet{tan12} preformed the test-particle simulation taking into account particle diffusion in the background fluid and synchrotron cooling,
and showed that the model reproduces the radial profiles of the photon index in 3C 58 and G21.5-0.9.
\citet{por16} also performed the test-particle simulation to obtain the diffusion coefficient based on the 3--D MHD simulation by \citet{2014MNRAS.438..278P}.
Their 1--D steady diffusion model with the obtained diffusion coefficient,
can explain the radial profile of the surface brightness and the photon index of 3C 58, G21.5-0.9 and the Vela.
The recent confirmation of the largely extended $\gamma$-ray halo around the Geminga pulsar by HAWC \citep{2017ApJ...843...40A}
also supports the idea of the particle diffusion \citep{2017Sci...358..911A}.
The extended gamma-ray would be emitted by electron--positron pairs diffusively escaped from the nebula.

However, the models including the diffusion effect are still in developing stage.
\citet{tan12} and \citet{por16}, which adopt the test-particle approximation, did not discuss the emission spectrum.
A model consistent with both the spectrum and radial profile is desired \citep{ishizaki17}.

In most of other astronomical phenomena, the spatial diffusion is considered for
energetically sub-dominant component with respect to the motion of the background fluid.
However, even energetically dominating particles 
are required to diffuse in order to reconcile the observed X-ray profile in PWNe.
Although we have a consensus that a diffusion coefficient of about $10^{27} \cmcms$ is required to reproduce the observed X-ray profile of PWNe \citep[e.g.,][]{tan12,por16},
this value implies that the energy flux carried by the diffusion is comparable to or larger than that carried by the advection with fluid.
In such cases, the test-particle approximation is not appropriate.

If we simply apply the particle diffusion fixing the velocity profile of the background fluid,
the energy and momentum conservations along the fluid are not assured.
We need to take into account the back reaction of the particle diffusion on the background fluid.
Furthermore, the idea of the particle diffusion in PWNe is ambiguous. The pulsar wind consists of only non-thermal particles,
which themselves are subject for the diffusion process,
so that the definition of the background fluid is not straightforward. Namely, the method to distinguish
the diffusive component from the background fluid is not apparent.

In this paper, we aim mainly to reproduce both the entire spectrum and X-ray radial profile of 3C 58 and G21.5-0.9 by improving the KC model
with the diffusion effect.
In Section \ref{sec:Model}, we explicitly define the background fluid 
and derive fluid equations satisfying the energy and momentum conservations with the diffusion effect.
In Section \ref{sec:dependence}, we demonstrate our 1--D steady state model
and discuss the velocity and magnetic field profiles in the flow, photon spectrum, and the radial profile of the surface brightness
especially for the model dependence on the diffusion coefficient.
In Section \ref{sec:application}, we apply our model to two observed sources 3C 58 and G21.5-0.9, which has been also discussed in \citet{ishizaki17}.
In Section \ref{sec:Discussion}, we summarize and discuss the results.

\section{Pulsar Wind Model with Diffusion}\label{sec:Model}

As we have mentioned, the concept of the diffusion in PWNe is ambiguous.
We cannot clearly distinguish the diffusive component from the background fluid.
This may indicate that the fluid approximation itself is not appropriate to deal with PWNe.
However, the concept of the advection with the fluid is still convenient to understand physics in PWNe.
In this section, retaining the fluid picture, we formulate the fluid equations
taking into account the spatial diffusion of particles self-consistently.
We solve those equations combined with the advection--diffusion equation for the energy distribution
function of non-thermal particles.

In Section \ref{sub:gen_form}, we clarify the definitions of the diffusion and background fluid,
then we present general formulation for the fluid equations describing the macroscopic motion of the non-thermal particles
and the transport equation describing the evolution of the energy spectrum of the particles.
In Section \ref{sub:1d}, we adopt those general equations to a spherically symmetric and steady system,
and show our model to apply for PWNe.
Note that, hereafter, the rest frame of the entire nebula is defined as reference frame $K$.

\subsection{Fluid equations and advection--diffusion equation}\label{sub:gen_form}

In order to treat the diffusion in a one-component fluid,
we assume a spatial diffusion coefficient with energy dependence,
with which the diffusion effect is negligible for the lowest energy particles,
which dominates in the number of particles.
The fluid velocity approximately corresponds to the average velocity of such low energy particles.
Note that energetically dominating particles, whose average energy is much higher than the lowest particle energy,
can be affected by the diffusion significantly.
Since the current for maintaining the magnetic field in the fluid is also dominated by low energy particles,
the frozen-in condition in the ideal MHD will be kept in this case.

The Boltzmann equation represented in the frame $K$ is written as
\begin{equation}\label{boltzmann}
\frac{\partial f}{\partial t}
+{\mbox{\boldmath $v$}}\cdot \nabla f
+ e\left[{\mbox{\boldmath $E$}}+\frac{{\mbox{\boldmath $v$}}}{c}\times{\mbox{\boldmath $B$}}\right]
\cdot\frac{\partial f}{\partial {\mbox{\boldmath $p$}}}
=S
\end{equation}
where $f$ is the phase space distribution functions of particles,
${\mbox{\boldmath $v$}}$ is the velocity of each particle,
${\mbox{\boldmath $E$}}$ and ${\mbox{\boldmath $B$}}$ are the electric field and magnetic field in the frame $K$,
and $S$ is the collision term.
The fluid flow velocity normalized by $c$ is obtained by averaging the velocity as
\begin{eqnarray}\label{def_beta_gen}
{\mbox{\boldmath $\beta$}}\equiv\frac{1}{c} \frac{\int {\mbox{\boldmath $v$}} f d^3p}{\int f d^3p},
\end{eqnarray}
and the corresponding Lorentz factor is $\gamma=1/\sqrt{1-\beta^2}$.
The co-moving frame of the fluid $K'$ is defined as the frame moving at velocity ${\mbox{\boldmath $\beta$}}$ with respect to the frame $K$.
Hereafter, the values in the frame $K'$ are expressed by primed characters, when it is necessary.
Even for the energetically dominating particles, the gyro radius is much shorter
than the typical fluid scale in PWNe (see discussion in \S \ref{sec:Discussion}).
This may allow us to assume that the momentum distribution is isotropic and the electric field is vanished in the frame $K'$, which
is equivalent to the ideal MHD condition,
\begin{eqnarray}\label{eq_gen:idealMHD}
{\mbox{\boldmath $E$}}+{\mbox{\boldmath $\beta$}}\times{\mbox{\boldmath $B$}}=0.
\label{eq:idealMHD}
\end{eqnarray}

Other than the macroscopic field in equation (\ref{eq:idealMHD}), a small scale turbulence of electromagnetic field may exist.
For simplicity, the collision term $S$, which expresses a stochastic process in a disturbed field,
is defined in the frame $K$ as
\begin{equation}\label{source}
S\equiv\nabla \cdot\left(\kappa \nabla f\right),
\end{equation}
where $\kappa$ is an energy-dependent diffusion coefficient\footnote{The diffusion process is usually defined in the fluid rest frame.
In a spherically symmetric steady flow, if we rewrite the collision term
with the coordinate of the frame $K$,
the diffusion coefficient for the radial direction is re-defined as an enlarged value by a factor of $\gamma^2$. The value $\kappa$
in equation (\ref{source}) is interpreted as such an effective value.}.
In general, the diffusion process is anisotropic. Since we will adopt the formulae in this section
to a spherically symmetric system, only the diffusion process toward radial direction is important.
So we calculate with the simple form of equation (\ref{source}).

In this paper, we do not solve equation (\ref{boltzmann}) directly.
First, we obtain fluid equations consistent with the Boltzmann equation (\ref{boltzmann}).
The energy and momentum equations in the fluid dynamics
corresponds to equations derived from integrating over the momentum space the products
of equation (\ref{boltzmann}) multiplied by energy $E_\pm$ or momentum ${\mbox{\boldmath $p$}}$, respectively.
The energy and momentum transfer due to the spatial diffusion
are obtained by the products from the term of equation (\ref{source}).
Adding those diffusion terms,
we obtain the energy conservation law,
\begin{multline}\label{eq_gen:energy}
\frac{1}{c}\frac{\partial}{\partial t}{}\left[\gamma^2\left(\epsilon+P\right)-P+\frac{E^2+B^2}{8\pi}\right]\\
+\nabla\cdot\left[
\gamma^2\left(\epsilon+P\right){\mbox{\boldmath $\beta$}}
+\frac{{\mbox{\boldmath $E$}}\times{\mbox{\boldmath $B$}}}{4\pi} \right.\\ \left.
-\nabla\left\{\left(\frac{4}{3}u^2+1\right)\delta\right\} 
+\left(\frac{4}{3}u^2+1\right){\mbox{\boldmath $\theta$}} 
\right]=-\gamma \Lambda,
\end{multline}
and the momentum conservation law,
\begin{multline}\label{eq_gen:momentum}
\frac{1}{c}\frac{\partial}{\partial t}{}\left[\gamma^2\left(\epsilon+P\right){\mbox{\boldmath $\beta$}}+\frac{{\mbox{\boldmath $E$}}\times{\mbox{\boldmath $B$}}}{4\pi}\right] \\
+\nabla\cdot\left[
	\gamma^2\left(\epsilon+P\right){\mbox{\boldmath $\beta$}}{\mbox{\boldmath $\beta$}}
	+P{\mbox{\boldmath $I$}}
	+\frac{{\mbox{\boldmath $E$}}{\mbox{\boldmath $E$}}+{\mbox{\boldmath $B$}}{\mbox{\boldmath $B$}}}{8\pi}  \right.\\ \left.
	-\nabla\left(\frac{4}{3}\gamma^2{\mbox{\boldmath $\beta$}}\delta\right)
	+\frac{4}{3}\gamma^2{\mbox{\boldmath $\beta$}}{\mbox{\boldmath $\theta$}}
\right]=-\gamma{\mbox{\boldmath $\beta$}} \Lambda,
\end{multline}
where $\epsilon$ and $P$ are energy density and pressure in the frame $K'$, respectively.
Hereafter, we assume that non-thermal particles are ultra-relativistic ($E'_\pm \simeq c p' \gg m_{\rm e} c^2$), which implies
\begin{eqnarray}\label{EOS}
P=\frac{1}{3}\epsilon.
\end{eqnarray}

The diffusion terms have been calculated with the fact that $f$ and $d^3{\mbox{\boldmath $p$}}/E_\pm$ are Lorentz invariant
and the assumption of the isotropic momentum distribution in the frame $K'$. The values,
$\delta$ and ${\mbox{\boldmath $\theta$}}$, in the diffusion terms
are defined by
\begin{eqnarray}\label{delta_gen}
\delta\equiv\frac{1}{c}\int \kappa E_\pm'f d^3{\mbox{\boldmath $p$}}',
\end{eqnarray}
\begin{eqnarray}\label{theta_gen}
{\mbox{\boldmath $\theta$}}\equiv\frac{1}{c}\int \left(\nabla{\kappa}\right) E_\pm'f d^3{\mbox{\boldmath $p$}}',
\end{eqnarray}
respectively.
The radiative cooling term is written with the cooling rate in the frame $K'$ as
\begin{eqnarray}\label{Q_gen}
\Lambda \equiv\frac{1}{c}\int Q_{\rm rad}' f d^3{\mbox{\boldmath $p$}}', 
\end{eqnarray}
where $Q'_{\rm rad}(E'_\pm)$ is the energy radiative loss rate per particle.
As shown in \citet{2010ApJ...715.1248T},
the radiative cooling by inverse Compton is negligibly inefficient compared to the synchrotron cooling so that
\begin{equation}
Q_{\rm rad}'= \frac{4}{3}\sigma_{\rm T}c\left(\frac{E_\pm'}{m_{\rm e}c^2}\right)^2\frac{B'^2}{8\pi}.
\end{equation}

The induction equation with the ideal MHD condition is
\begin{eqnarray}\label{eq_gen:induction_eq}
\frac{1}{c}\frac{\partial {\mbox{\boldmath $B$}}}{\partial t}=\nabla\times\left({\mbox{\boldmath $\beta$}}\times{\mbox{\boldmath $B$}}\right).
\end{eqnarray}
If $\delta$, ${\mbox{\boldmath $\theta$}}$, and $\Lambda$ are given,
the fluid quantities ${\mbox{\boldmath $\beta$}}$, $\epsilon$, p, ${\mbox{\boldmath $E$}}$ and ${\mbox{\boldmath $B$}}$ can be calculated
by solving equations (\ref{eq_gen:energy}), (\ref{eq_gen:momentum}), (\ref{EOS}), (\ref{eq_gen:induction_eq}) and (\ref{eq_gen:idealMHD}) (hereafter fluid equations).
However, those quantities depending on the functional form of $f$ are not given in advance.

In order to obtain $\delta$, ${\mbox{\boldmath $\theta$}}$, and $\Lambda$,
we also solve the advection--diffusion equation \citep[hereafter AD equation, e.g.,][]{1964ocr..book.....G}
written as
\begin{multline}
\frac{\partial n({\mbox{\boldmath $x$}},E_\pm')}{\partial t}+
\nabla\cdot\left[c{\mbox{\boldmath $u$}}n({\mbox{\boldmath $x$}},E_\pm')-\kappa\nabla n({\mbox{\boldmath $x$}},E_\pm')\right]\\
-\frac{\partial}{\partial E_\pm'}\left[Q' n({\mbox{\boldmath $x$}},E_\pm')\right]
=0,
\label{eq_gen:FPeq}
\end{multline}
where $n({\mbox{\boldmath $x$}},E_\pm')=4\pi p'^2f/c$ is the energy spectrum of particles in the frame $K'$,
${\mbox{\boldmath $u$}}=\gamma {\mbox{\boldmath $\beta$}}$ is the flow 4-velocity, and $Q'$ is the energy loss rate
due to the radiative and adiabatic cooling per particle.
This equation can be solved if ${\mbox{\boldmath $u$}}$ and $B'$ are given in advance,
so that the fluid equations and the AD equation are complementary each other.
We can solve the AD equation and the fluid equations alternately until the energy densities
estimated from both the methods agree with each other.

\subsection{Spherical steady nebulae}\label{sub:1d}

We adopt the fluid equations and the AD equation to the spherical steady outflow,
and assume that the magnetic field is purely toroidal as the KC model assumed.
The magnetic field in the frame $K'$ is $B'=B/\gamma$,
and the induction equation (\ref{eq_gen:induction_eq}) leads to
\begin{equation}
r\beta B=\const .
\label{eq:induction_eq_1d}
\end{equation}
The diffusion coefficient is assumed to be expressed by a separable form as
\begin{equation}
\kappa(r,E'_\pm)=\phi(r)\tilde{\kappa}(E'_\pm).
\label{eq:separable}
\end{equation}
Then, the terms with $\delta$ and ${\mbox{\boldmath $\theta$}}$ in the fluid equations can be put together.
Equations (\ref{eq_gen:energy}) and (\ref{eq_gen:momentum}) are written as
\begin{multline}\label{1d_wind_beq_ene}
\frac{1}{r^2}\frac{\partial}{\partial r}{}\left[
r^2\left(
\gamma^2\left(\epsilon+P\right)\beta
+\frac{B^2}{4\pi}\beta  \right.\right.\\\left.\left.
-\phi(r)\frac{\partial}{\partial r}{}\left\{\left(\frac{4}{3}\gamma^2\beta^2+1\right)\tilde{\delta}(r)\right\}
\right)
\right]=-\gamma \Lambda(r),
\end{multline}
and
\begin{multline}\label{1d_wind_beq_mom}
\frac{1}{r^2}\frac{\partial}{\partial r}{}\left[
r^2\left(
\gamma^2\left(\epsilon+P\right)\beta^2
+P
+\frac{B^2}{8\pi}\left(1+\beta^2\right)  \right.\right.\\ \left.\left.
-\phi(r)\frac{\partial}{\partial r}{}\left\{\frac{4}{3}\gamma^2\beta\tilde{\delta}(r)\right\}
\right)
\right]=\frac{2P}{r}-\gamma\beta \Lambda(r),
\end{multline}
respectively, where $\tilde{\delta}$ is defined as
\begin{eqnarray}\label{1d_wind_beq_delta_tilde}
\tilde{\delta}(r)\equiv\frac{1}{c}\int \tilde{\kappa} E_\pm'n(r,E_\pm') dE_\pm' .
\end{eqnarray}
The AD equation (\ref{eq_gen:FPeq}) in a spherical steady system becomes
\begin{multline}
\frac{1}{r^2}\frac{\partial}{\partial r}\left[r^2\left(cu(r)n(r,E_\pm')-\kappa\frac{\partial n(r,E_\pm')}{\partial r}\right)\right]\\
-\frac{\partial}{\partial E_\pm'}\left[Q' n(r,E_\pm') \right]
=0,
\label{eq:FP_1d}
\end{multline}
where $Q'$ expresses the radiative and adiabatic cooling effects as
\begin{equation}
Q'=Q'_{\rm rad}+\frac{E_\pm'}{3}\frac{c}{r^2}\frac{\partial }{\partial r}\left[r^2 u(r)\right].
\end{equation}

We solve only the downstream of the termination shock at $r=r_{\rm s}$.
Given $r_{\rm s}$, the energy release rate $L_{\rm sd}$, and the magnetization parameter,
\begin{equation} 
\sigma\equiv\frac{B_{\rm u}^2/4\pi}{n_{\rm u}u_{\rm u}\gamma_{\rm u}m_{\rm e} c^2},
\end{equation}
where $B_{\rm u}$, $n_{\rm u}$, and $\gamma_{\rm u}=\sqrt{1+u_{\rm u}^2}$ are $B$, total particle density,
and $\gamma$ at just upstream of the shock, respectively,
the Rankine--Hugoniot jump condition \citep[for details see][]{1984ApJ...283..694K} provides the boundary condition
at $r=r_{\rm s}$ for the fluid equations.

As the inner boundary condition for the AD equation, we assume a broken power-law energy distribution,
\begin{eqnarray} 
n(r_{\rm s},E_\pm')=\left\{
\begin{array}{ll}
\displaystyle\frac{n_0}{E_{\rm b}}\left(\frac{E_\pm'}{E_{\rm b}}\right)^{-p_1} & \left( E_{\rm min}<E_\pm'<E_{\rm b} \right) \\
\displaystyle\frac{n_0}{E_{\rm b}}\left(\frac{E_\pm'}{E_{\rm b}}\right)^{-p_2} & \left( E_{\rm b}<E_\pm'<E_{\rm max} \right) 
\end{array}
\right.,
\label{eq:bPL}
\end{eqnarray}
where $n_0$ is a normalization factor which is adjusted to be consistent with the boundary condition of the fluid equations,
$E_{\rm b}$ is the intrinsic break energy,
$E_{\rm min}$ and $E_{\rm max}$ are the minimum and maximum energy, respectively,
and $p_1$ and $p_2$ are power-law indices for low- and high-energy potion of the particle spectrum, respectively.
Hereafter, we also assume $p_1<2<p_2$, which implies that
particles with energy $\sim E_{\rm min}$ dominate in the number of particles, and particles with energy $\sim E_{\rm b}$ dominate in the energy density of particles.

The density at the edge of the nebula depends on the flow velocity and diffusion coefficient outside the nebula, which are not considered in this paper.
If the diffusion coefficient outside the nebula is very smaller than the inner value,
particles pile up around the edge of the nebula.
However, the diffusion coefficient in the ISM is larger than the value adopted in this paper,
so that the pile-up case may be unlikely.
Here we assume that the contribution of the re-entering particles from ISM/SNR is not so large.
As the simplest case, we take an outer boundary condition,
\begin{equation}\label{Boundary}
\frac{\partial^2}{\partial r^2}\left[r^2 n(r,E'_\pm)\right]_{r=r_{\rm N}}=0.
\end{equation}
This condition makes the density profile smoothly connect to the outside profile of the steady solution without advection, $n(r,E'_{\pm})\propto r^{-1}$.

The diffusion process in PWNe is highly uncertain.
For simplicity, we assume that the energy dependence of $\kappa$
is Kolmogorov-like as
\begin{equation}
\tilde{\kappa}(E_\pm)=\kappa_0\left(\frac{E_\pm}{E_{\rm b}}\right)^{1/3},
\label{eq:kappa_tild}
\end{equation}
where the parameter $\kappa_0$ is constant.

As for the spatial dependence of $\kappa$,
the simplest model is the homogeneous diffusion with $\phi(r)=1$.
However, in such a case, the diffusion can influence the shock jump condition.
The diffusion effect on the termination shock is beyond the scope of this paper.
To avoid a complicated situation with a modified jump condition by diffusion,
we neglect the effect of the diffusion near the termination shock.
We turn on the diffusion effect at a certain radius $r_{\rm diff}>r_{\rm s}$.
However, a sudden onset of the diffusion effect induces numerical instability.
We assume an artificial function form as
\begin{equation}
\phi(r)=\frac{1}{2}\left[1+\tanh\left(\frac{r-r_{\rm diff}}{\Delta r}\right)\right],
\label{eq:kappa_fr}
\end{equation}
where $\Delta r$ is a transition scale of the smoothing.
This function smoothly changes from zero (for $r_{\rm diff}-r \gg \Delta r$) to unity (for $r - r_{\rm diff} \gg \Delta r$).
The equation (\ref{eq:kappa_fr}) is introduced just to avoid the technical issue in numerical calculation.
As we will see later,
the diffusion effect in the inner part of the nebula is negligible.
As long as we take significantly small $r_{\rm diff}$ and $\Delta r$,
the diffusion can be regarded as almost homogeneous,
and the result does not depend on the details of this artificial functional form.
The very weak dependence on $r_{\rm diff}$ and $\Delta r$ in our results has been checked.

The parameters other than the diffusion coefficient are the same as those in \citet{ishizaki17}.
Hereafter, we assume that the minimum energy is fixed as $E_{\rm min}=10m_{\rm e}c^2$,
and that the maximum energy $E_{\rm max}$ is determined as the energy at which a gyro radius is equal to the shock radius.
The nebula size $r_{\rm N}$ and the energy release rate, which is the same as the spin down luminosity $L_{\rm sd}$, are obtained from observation.
In summary, the parameters to be adjusted are six: $r_{\rm s}$, $\sigma$, $E_{\rm b}$, $p_1$, $p_2$ and $\kappa_0$.

Since the AD equation is 2-dimensional elliptic equation,
it is solved using the finite difference method and the SOR method \citep[e.g.,][]{1992nrfa.book.....P}.
The fluid equations are integrated using a 4-th order Runge--Kutta method.
Until the outputs from the fluid equations and the AD equation become consistent with each other,
we iterate the calculation.
All of results shown in this paper satisfy an accuracy of $\mathcal{O}(1)$ \% in the energy/momentum conservation.

\section{Diffusion Effects}\label{sec:dependence}

\begin{figure*}[!tb]
	\begin{center}
		$\begin{array}{cc}
		\includegraphics[width=0.5\textwidth]{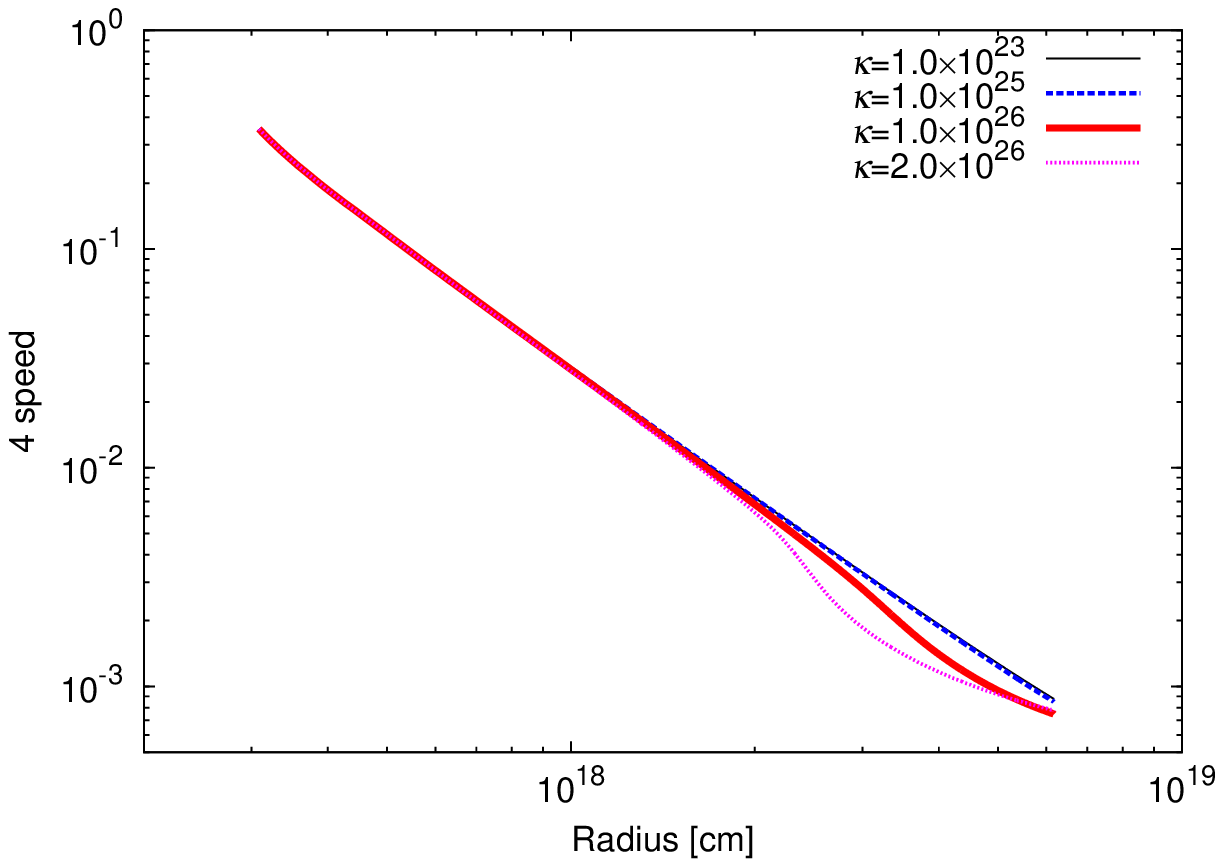} &
		\includegraphics[width=0.5\textwidth]{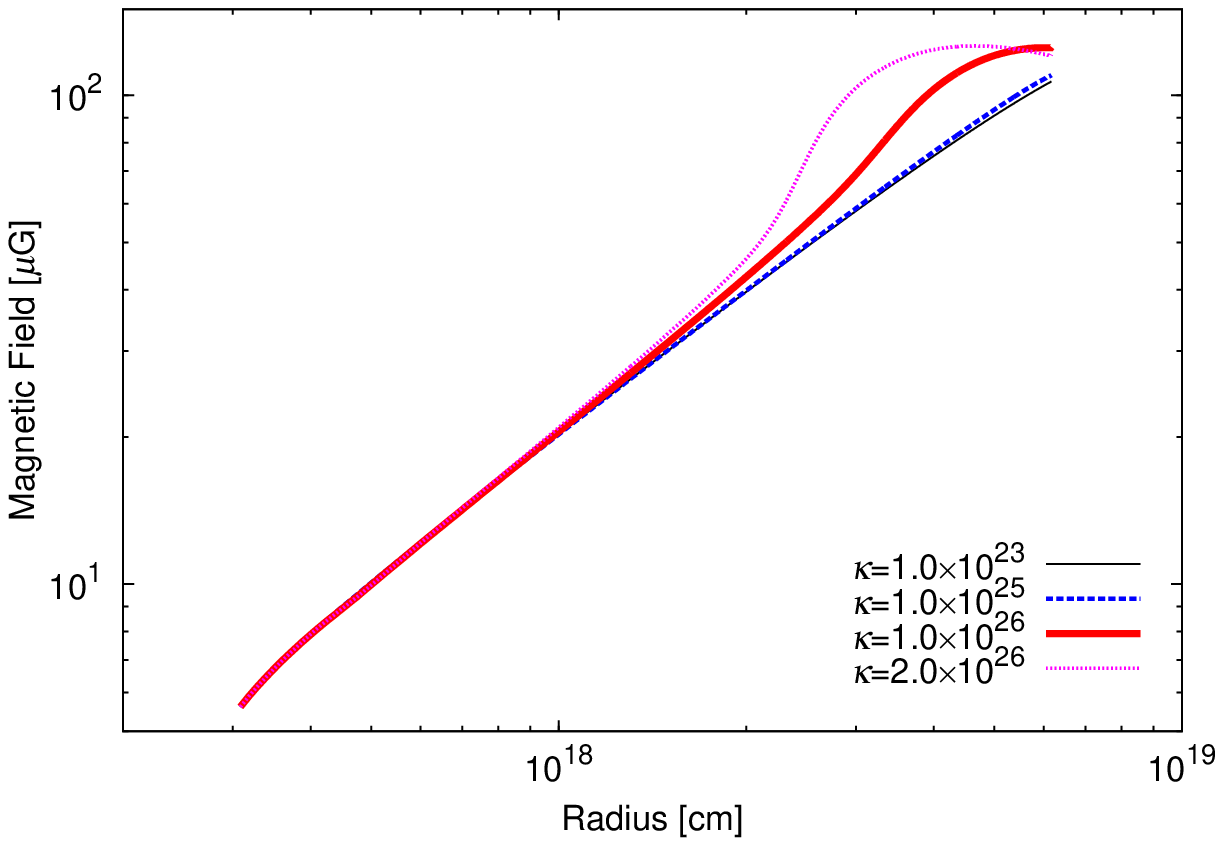}
		\end{array}$
	\end{center}
	\caption{
		Radial profiles of 4-speed (Left) and magnetic field (Right) in the test calculations for various $\kappa_0$.
	}
	\label{demo:fluid}
\end{figure*}

In this section,
we show example results with various values of $\kappa_0$,
and fixed values of the other parameters ($L_{\rm sd}$, $E_{\rm b}$, $p_1$, $p_2$, $r_{\rm s}$, $r_{\rm N}$ and $\sigma$).
Here, to help understand the dependence on the $\kappa_0$,
we define the length scale $r_{\rm pe}$ where the diffusion process begins to become effective
for the energetically dominating particles, whose energy is $E_{\rm b}$, by
\begin{equation}\label{pecle}
r_{\rm pe}\equiv\frac{r_{\rm s}^2cu_{\rm d}}{2\kappa_0}\sim 5.0\times10^{18}
\left(\frac{r_{\rm s}}{0.1 \pc}\right)^2 \left(\frac{\kappa_0}{10^{26} \cmcms}\right)^{-1} \cm,
\end{equation}
where $u_{\rm d}$ is the four speed of the flow at just downstream of the termination shock
and approximately equal to $1/\sqrt{8}$ for $\sigma\ll 1$.
This radius corresponds to the location where the advection and diffusion timescales are the same
\footnote{
	This is obtained from the analytical steady state solution of advection diffusion equation
	assuming that the flow is incompressible, the diffusion coefficient is uniform in space, and the cooling process is negligible.}.
In this section, the model of the inter stellar photon field is the same as for G21.5-0.9 \citep[see][]{ishizaki17}.
This can be approximated by mainly three thermal components as follows:
(1) CMB: energy density $0.25 \ {\rm eV/cc}$ and temperature $2.7 \ {\rm K}$,
(2) dust emission: energy density $0.65 \ {\rm eV/cc}$ and temperature $35 \ {\rm K}$,
(3) stellar emission: energy density $2.0 \ {\rm eV/cc}$ and temperature $4000 \ {\rm K}$.
As an example, the parameter values other than $\kappa_0$ are set as
$L_{\rm sd}=10^{38}\ {\rm erg}~{\rm s}^{-1}$, $E_{\rm b}=10^5 m_{\rm e} c^2$, $p_1=1.1$, $p_2=2.5$,
$r_{\rm s}=0.1 \pc$, $r_{\rm N}=2.0 \pc$ and $\sigma=10^{-4}$.
We test changing $\kappa_0$ as $10^{23}$, $10^{25}$, $10^{26}$, and $2\times 10^{26} \cmcms$.
We also set the parameters in equation (25) as $r_{\rm diff}=0.14\pc$ and $\Delta r=0.001\pc$.

\begin{figure}[!tb]
	\centering
	\includegraphics[width=1.0\linewidth]{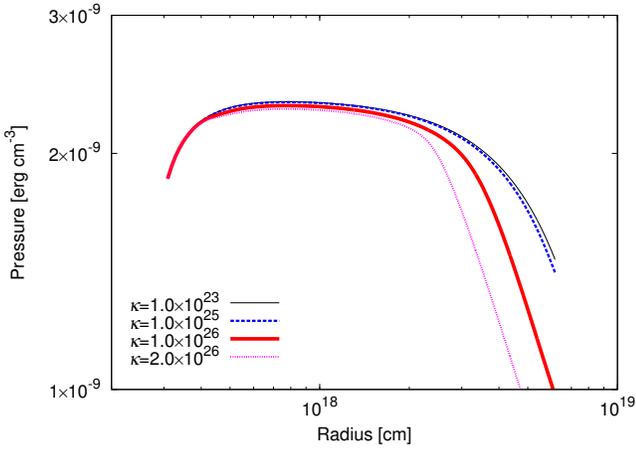}
	\caption[test]{
		Radial profiles of pressure in the test calculations for various $\kappa_0$.
	}
	\label{demo:pressure}
\end{figure}

The case of $\kappa_0=10^{23} \cmcms$ is almost the same as the KC model,
in which the diffusion is neglected ($r_{\rm pe} \gg r_{\rm N}$).
Figure \ref{demo:fluid} shows that the wind is decelerated at outer region as $\kappa_0$ increases.
Accordingly, the magnetic field is amplified conserving the value of $r \beta B$.
Let us see the case of $\kappa_0=10^{26} \cmcms$ as the benchmark case
(red thick lines in Figure \ref{demo:fluid}).
The deviation from the lines for low $\kappa_0$ models
is seen around $r=3\times10^{18} \cm$.
The radius of $r_{\rm pe}$ is consistent with this deceleration radius within a factor 2.
As seen in Figure \ref{demo:pressure},
at the deceleration radius the pressure also starts deviating from the KC model.
Outside $r_{\rm pe}$ the pressure decreases steeper than the pressure profile in the KC model.
In spherically symmetric system,
the particles basically diffuse outward,
so that the diffusion flux brings out the fluid momentum outward.
Since the fluid receives inward reaction force via particle diffusion, the fluid starts decelerating at $r=r_{\rm pe}$.

\begin{figure}[!tb]
	\centering
	\includegraphics[width=1.0\linewidth]{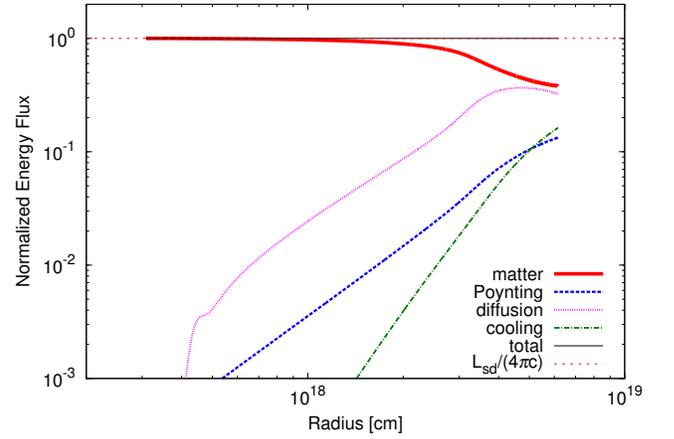}
	\caption[test]{
		The radial profiles of the energy flow outward through the area of $4\pi r^2$ per unit time normalized by the value at the $r=r_{\rm s}$
		for the case of $\kappa_0=10^{26} {\rm cm^2}~{\rm s^{-1}}$.
		The line labeled "cooling" is the total amount of synchrotron radiation per unit time emitted inside the radius $r$ divided by $4\pi c$.
	}
	\label{demo:energy}
\end{figure}

After the sudden deceleration at $r \sim r_{\rm pe}$, the deceleration is rather gradual.
As shown in Figure \ref{demo:energy}, far outside $r_{\rm pe}$,
the energy fluxes due to diffusion and advection are balanced, so that
\begin{equation}
\gamma^2\left(\epsilon+P\right)\beta \sim
\frac{\partial}{\partial r}\left[\left(\frac{4}{3}\gamma^2\beta^2+1\right)\tilde{\delta}\right] .
\end{equation}
In the limit of $\beta \ll1$ and approximation of $\delta\sim\kappa\epsilon$,
we obtain $\beta\propto r^{-1}$ and $B\propto r^{0}$.
This simple estimate is consistent with the behavior in our calculation result.

\begin{figure}[!tb]
	\centering
	\includegraphics[width=1.0\linewidth]{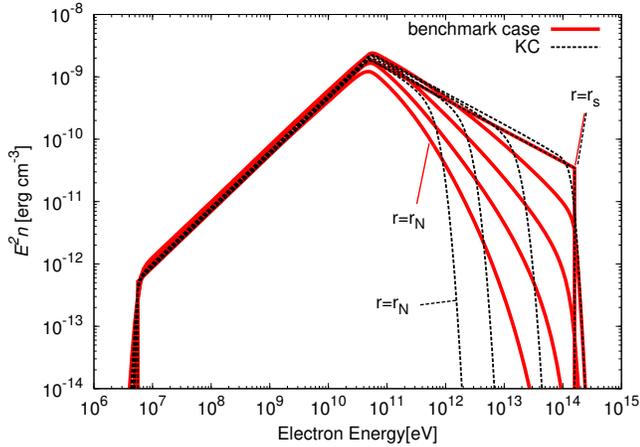}
	\caption[test]{
		The energy spectrum of pairs at $r=r_{\rm s}$, $5r_{\rm s}$, $10r_{\rm s}$, $15r_{\rm s}$, and  $r_{\rm N}=20r_{\rm s}$.
		Red solid lines represent the model taking into account the diffusion, and black dashed line represent the KC model.
		The parameters are same as Fig. \ref{demo:energy}.
	}
	\label{demo:electron}
\end{figure}

Figure \ref{demo:electron} shows the evolution of the particle energy spectrum for the benchmark case
comparing with the KC model.
As we have expected, in the benchmark case, the spectral shapes show that
the low-energy particles with $E'_\pm \sim E_{\rm min}$ are not affected by the diffusion,
which modifies the spectrum above $E'_\pm \sim E_{\rm b}$.
In the KC model, the sharp cooling cutoff in the spectrum shifts to lower energy with radius.
In contrast, in the model with diffusion, the energy spectrum does not show a sharp cutoff.
The energy dependence of the diffusion time vanishes the sharp cutoff feature.
As shown in the line labeled $r=r_{\rm N}$ in Figure \ref{demo:electron},
particles with $E\sim10^{13} \eV$, which are main population responsible for emitting X-rays,
exist in the benchmark case, while such particles can not exist at $r=r_{\rm N}$ in the KC model.
In our calculation, 
particles with higher energy escape through the boundary $r=r_{\rm N}$ more efficiently via the diffusion.
As a result, given the radius $r$, the spectrum with the diffusion effect
becomes softer than the KC model spectrum.
This behavior is similar to the result of the time-dependent one-zone model by \citet{2012MNRAS.427..415M},
which takes into account the diffusive escape from the nebula.
The spectral softening due to particle diffusion is the same mechanism as that seen in the cosmic-ray
spectrum in our galactic plane \citep[e.g. see][]{1998ApJ...509..212S}.

The amplification of the magnetic field by the back-reaction of the diffusion also affects the particle spectrum through the synchrotron cooling.
As seen in Figure \ref{demo:fluid}, in the benchmark case, the magnetic field is stronger than the KC model at any radius.
As the diffusion is more efficient for particles with higher energy,
even if they are affected by the strong radiative cooling,
such particles are able to spread to outer side of the nebula.

\begin{figure}[!tb]
	\centering
	\includegraphics[width=1.0\linewidth]{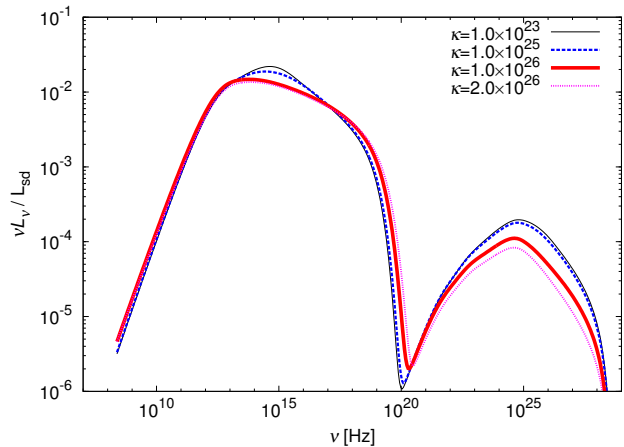}
	\caption[test]{
		The entire photon spectrum calculated for various values of $\kappa_0$.
	}
	\label{demo:Luminosity}
\end{figure}

Based on the particle spectra shown in Figure \ref{demo:electron} and the magnetic field shown in Figure \ref{demo:fluid},
we can calculate the emission from the nebula.
For the calculation of the emission, we adopt the same method as \citet{ishizaki17}.
We consider only synchrotron radiation and ICS including the Klein--Nishina effect.
The model of the interstellar radiation field is taken from GALPROP v54.1 \citep[][and references therein]{vla11},
in which the results of \citet{por05} are adopted.

Figure \ref{demo:Luminosity} shows the entire spectra of the nebula for each $\kappa_0$.
In the case of $ \kappa_0=10^{23} \cmcms$, of course, the spectrum is almost the same as that in the KC model.
The entire spectrum of the benchmark case becomes harder than the KC model.
As seen in Figure \ref{demo:electron}, the particle energy spectrum becomes softer with increasing radius.
However, since the high energy particles are diffused to outside region, where the magnetic field is strong,
high-energy synchrotron photons are efficiently emitted from the outer region.
As a result, this mechanism hardens the integrated spectrum.
On the other hand, the ICS flux monotonically decreases with increasing $\kappa_0$.
This is because particles escape efficiently from the nebula for a high $\kappa_0$.
Here, we have neglected the emission outside the nebula so that the ICS flux decreases with $\kappa_0$.

\begin{figure}[!tb]
	\centering
	\includegraphics[width=1.0\linewidth]{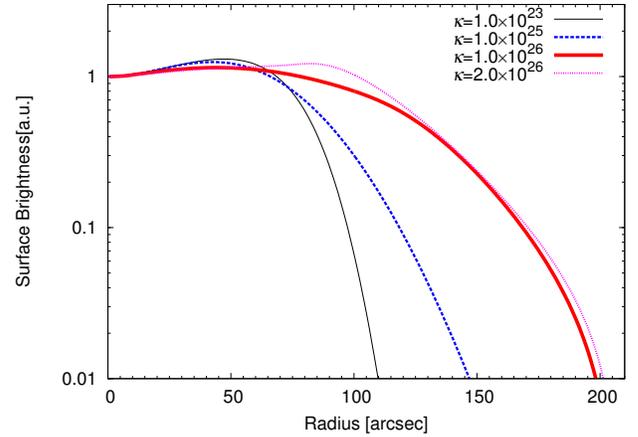}
	\caption[test]{
		The dependence on the $\kappa_0$ of the X-ray surface brightness profile for 0.5-10 keV range. 
		Here we assume that the distance to the nebula is $2$ kpc.
		The edge of the nebula $r_{\rm N}$ is corresponding to $200 ''$.
	}
	\label{demo:brightness}
\end{figure}

\begin{figure}[!tb]
	\centering
	\includegraphics[width=1.0\linewidth]{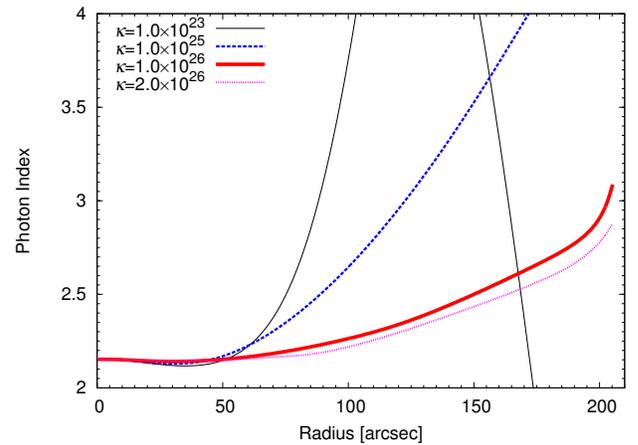}
	\caption[test]{
		The dependence on the $\kappa_0$ of the photon index profile for 0.5-10 keV range.
		The distance and the nebula size are the same as those in Fig. \ref{demo:brightness}
	}
	\label{demo:photon_index}
\end{figure}

In Figure \ref{demo:brightness}, the X-ray surface brightness profile is shown.
As expected, the size of X-ray emission region becomes larger with increasing $\kappa_0$.
Although the back reaction by diffusion on the fluid dynamics
is negligible for $\kappa_0=10^{25} \cmcms$ (see Figure \ref{demo:fluid}),
the X-ray size is significantly enlarged compared to the KC model.
In the benchmark case, the emission region of X-rays extends to the edge of the nebula, where the magnetic field is strong.
The peak radius in X-ray band roughly corresponds to the radius at which the magnetic field amplification is saturated.
Outside the peak radius, the radiative cooling becomes efficient.
The radial profile of the photon index in X-rays is shown in Figure \ref{demo:photon_index}.
For all the cases, the photon index shows a softening with increasing radius.
Moreover, the photon index is kept harder as $\kappa_0$ is larger.
The sudden softening at a certain radius in the KC model \citep{Rey03,2004ApJ...616..403S}
is not seen in our diffusion models.

\begin{figure*}[!tb]
	\begin{center}
		$\begin{array}{cc}
		\includegraphics[width=0.5\textwidth]{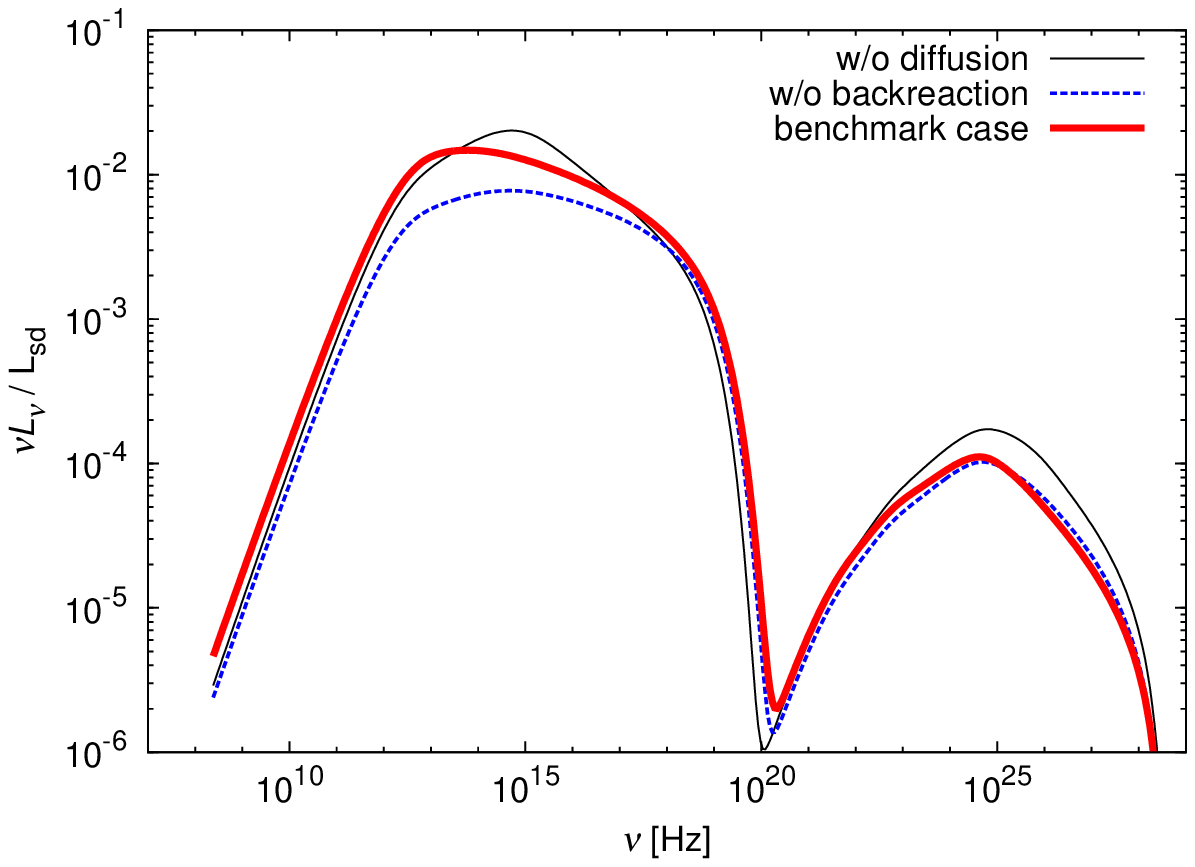} &
		\includegraphics[width=0.5\textwidth]{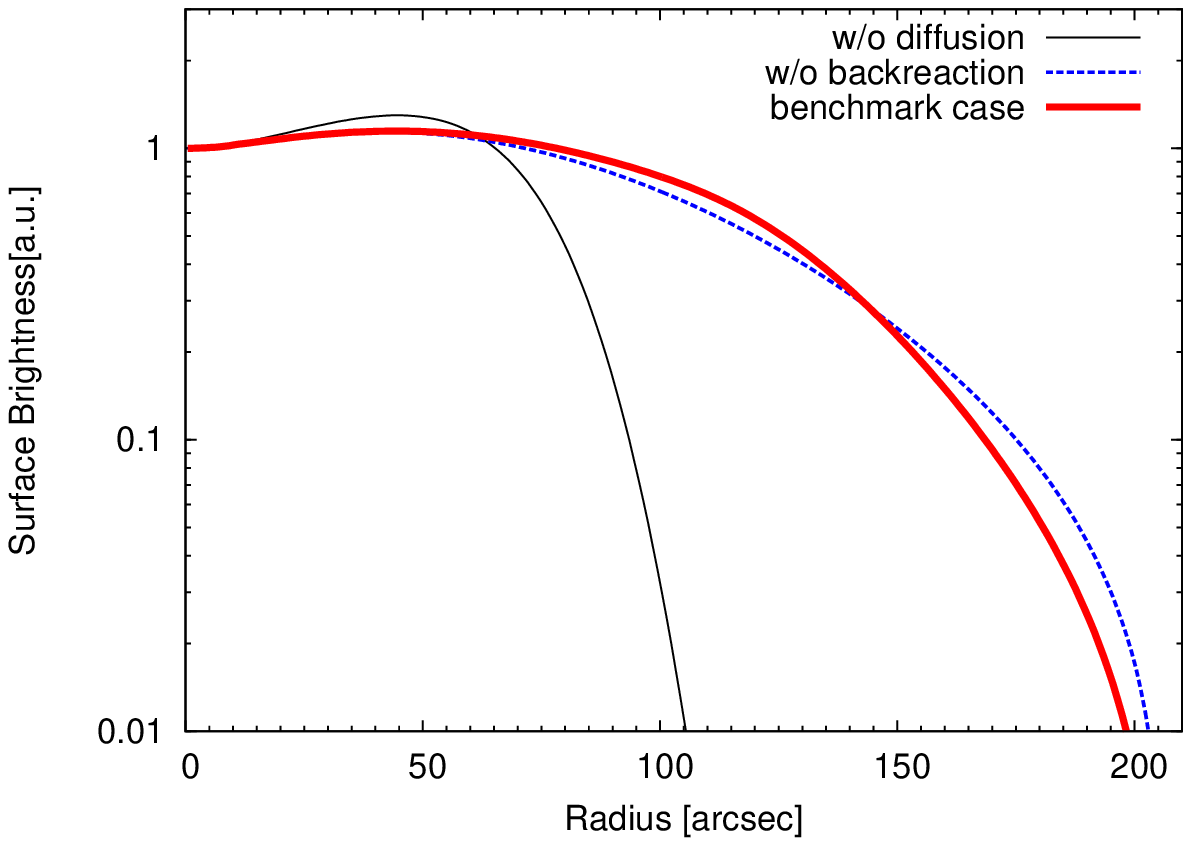}
		\end{array}$
	\end{center}
	\caption{
		Comparison of the self-consistent model (benchmark model, red thick), the diffusion model without the back reaction (blue dashed), and the KC model (black thin).
		(Left) The entire spectra. (Right) The X-ray surface brightness for 0.5-10 keV range.
The diffusion coefficient is commonly $\kappa_0=10^{26} \cmcms$ for
the self-consistent model and the model without the back reaction.
	}
	\label{demo:backreaction}
\end{figure*}

Finally, to check how much the back reaction of diffusion affects the nebula emission,
we compare the calculations with and without the back reaction of diffusion to the fluid.
The model without the back reaction is obtained solving the AD equation
with $\kappa_0=10^{26} \cmcms$ for the given velocity field in the KC model.
In this case, the energy and momentum conservations are not ensured.
In Figure 8, the entire spectrum and the X-ray surface brightness are compared with those of the benchmark model and KC model ($\kappa_0=0$).
In the model without the back reaction,
the magnetic field is not amplified,
so that the synchrotron flux becomes half of the flux of the benchmark model.
Furthermore, since the flow velocity is decelerated in the benchmark case,
the advection time
\begin{equation}
	t_{\rm adv}=\int_{r_{\rm s}}^{r_{\rm N}}\frac{dr}{cu(r)},
	\label{eq:def_tadv}
\end{equation}
in the benchmark case is longer than that in the KC model.
Due to the stronger magnetic field and the longer advection time,
the spectral peak, which corresponds to the cooling break, appears at lower frequency than the case without the back reaction.
On the other hand, the X-ray sizes in the benchmark model and the model without the back reaction are not much different.

\section{Application to 3C 58 and G21.5-0.9}\label{sec:application}

In this section, we apply our model to the two PWNe 3C 58 and G21.5-0.9, for which we have rich observational data especially in X-rays.
Observational properties of these PWNe are summarized in Table \ref{BestFitParameter} \citep[for details see][]{ishizaki17}.
In these objects,
for parameter sets that reproduce the entire spectra,
the extent of X-rays in the KC model inconsistently becomes smaller than the observed extent \citep{ishizaki17}.

As \citet{ishizaki17} discussed,
the spectral indices of the observed spectra in X-rays can not be reproduced by the KC model.
If parameters are adjusted to reconcile the X-ray spectral index,
the IR/Opt and radio emission can not be reproduced.
Below we discuss whether the diffusion effects can resolve those problems or not.
Hereafter we call the model with diffusion the DF model.

\begin{table*}[!tbp]
	\caption{Parameters in our calculations.}
	\begin{center}
		\scalebox{0.8}{
			\begin{tabular}{lcrrrr}
				\hline\hline
				&  & \multicolumn{2}{c}{3C 58} & \multicolumn{2}{c}{G21.5-0.9} \\ 
				Given Parameters & Symbol & \multicolumn{1}{c}{KC$^{\rm a}$} & \multicolumn{1}{c}{DF} & \multicolumn{1}{c}{KC$^{\rm a}$} & \multicolumn{1}{c}{DF} \\ 
				\hline
				Spin-down Luminosity {\rm (erg~s$^{-1}$)} & $L_{\rm sd}$ & \multicolumn{2}{c}{$3.0\times10^{37}$} & \multicolumn{2}{c}{$3.5\times10^{37}$}\\ 
				Distance (kpc) & $D$ & \multicolumn{2}{c}{$2.0^{\rm b}$} &\multicolumn{2}{c}{$4.8^{\rm c}$} \\ 
				Radius of the nebula (pc) & $r_{\rm N}$ &\multicolumn{2}{c}{2.0} & \multicolumn{2}{c}{0.9}  \\ 
				\hline
				Fitting Parameters &  &  &  &  &  \\ 
				\hline
				Break Energy (eV) & $E_{\rm b}$ & $4.1\times10^{10}$ & $3.0\times10^{10}$ &  $2.6\times10^{10}$ & $6.0\times10^{10}$ \\ 
				Low-energy power-law index & $p_1$ & 1.26 & 1.08 & 1.1 & 1.2  \\ 
				High-energy power-law index & $p_2$ & 3.0 & 2.9 & 2.3 & 2.5 \\ 
				Radius of the termination shock (pc) & $r_{\rm s}$ & 0.13 & 0.14 & 0.05 & 0.05  \\ 
				Magnetization parameter & $\sigma$ & $1.0\times10^{-4}$ & $2.0\times10^{-4}$ & $2.0\times10^{-4}$ & $6.0\times10^{-4}$\\ 
				Diffusion coefficient at $E_{\rm b}$ (cm$^2$~s$^{-1}$) & $\kappa_0$ & - & $1.0\times10^{26}$ & - & $1.0\times10^{26}$  \\
				\hline
				Obtained Parameters &  &  &  &  &  \\ 
				\hline
				Initial bulk Lorentz factor & $\gamma_{\rm u}$ & $7.3\times10^{3}$ & $2.4\times10^{4}$ & $2.1\times10^{4}$ & $1.9\times10^{4}$ \\ 
				Pre-shock density (cm$^{-3}$) & $n_{\rm u}$ & $1.1\times10^{-11}$ & $2.7\times10^{-12}$ & $1.1\times10^{-11}$ & $1.3\times10^{-11}$ \\ 
				Pre-shock magnetic field ($\mu$G) & $B_{\rm u}$ & 0.79 & 1.0 & 3.1 & 5.4 \\ 
				Maximum energy (eV) & $E_{\rm max}$ & $9.5\times10^{13}$ & $1.3\times10^{14}$ & $1.4\times10^{14}$ & $2.5\times10^{14}$ \\ 
				Average magnetic field ($\mu$G) & $B_{\rm ave}$ & 31 & 34 & 120 & 133 \\ 
				Advection time (year) & $t_{\rm adv}$ & 1500 & 1300 & 800 & 630 \\ 
				Flow velocity at $r=r_{\rm N}$ (km s$^{-1}$) &  & 490 & 540 & 460 & 720 \\ 
				Ratio of $r_{\rm pe}$ to $r_{\rm N}$ & $r_{\rm pe}/r_{\rm N}$ & - & 3.2 & - & 0.91  \\
				\hline\hline
			\end{tabular}
		}
	\end{center}
	{\scriptsize 
		$^{\rm a}$ \citet{ishizaki17}; $^{\rm b}$ \citet{2013A&A...560A..18K}; $^{\rm c}$ \citet{2008MNRAS.391L..54T}.
	}
	\label{BestFitParameter}
\end{table*}

The model parameters and results are summarized in Table \ref{BestFitParameter}.
For comparison, we also show the result of ``broadband model'' in \citet{ishizaki17} (denoted "KC" in Table \ref{BestFitParameter}),
in which the diffusion effect is not incorporated.
For each object,
despite the fact that the wind is more decelerating in the DF model,
the advection time in the DF model is shorter than that in the KC model.
This is because $\sigma$ in the DF model is larger than that of the KC model.
We also tabulate the ratio of $r_{\rm pe}$ to $r_{\rm N}$,
which can be used as a criterion for the efficiency of diffusion.
For 3C 58, $r_{\rm pe}/r_{\rm N} > 1$, so that the diffusion has not much effect on the fluid motion.
In contrast, in the case of G21.5-0.9, $r_{\rm pe}/r_{\rm N}$ is smaller than unity,
and this indicates that diffusion is efficient with this parameter set.
For the DF models, we adjusted the parameters in equation (\ref{eq:kappa_fr}) as $|r_{\rm diff}-r_{\rm s}| \leq 0.03 r_{\rm N}$ and $\Delta r \leq 0.008 r_{\rm N}$.
Those small values may not affect the results largely.

\begin{figure*}[!tb]
	\begin{center}
		$\begin{array}{cc}
		\includegraphics[width=0.5\textwidth]{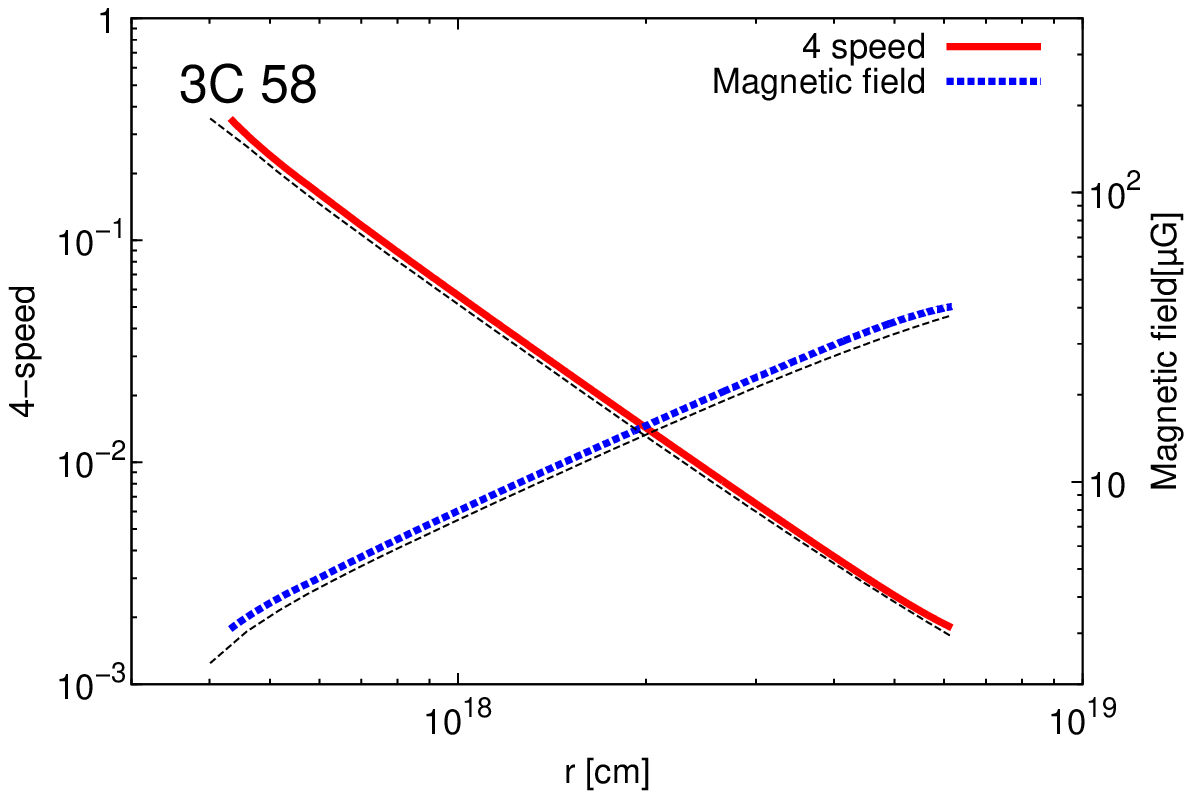} &
		\includegraphics[width=0.5\textwidth]{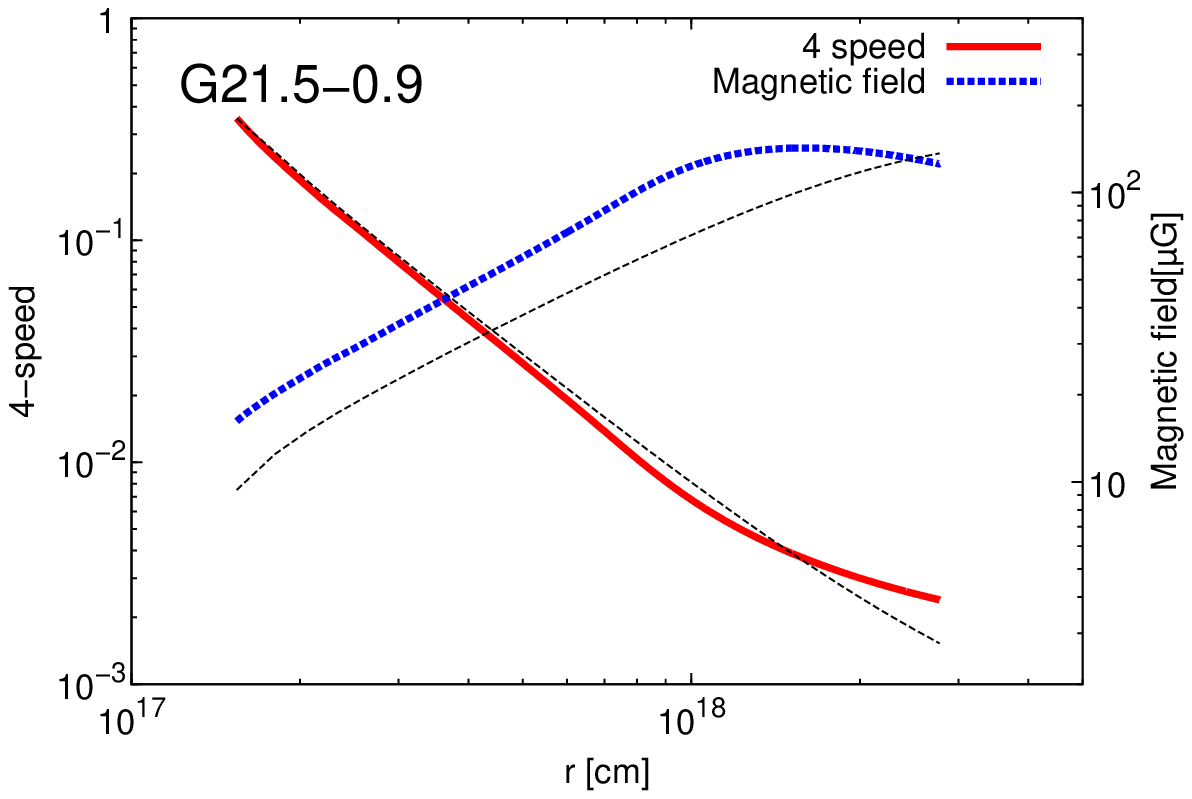}
		\end{array}$
	\end{center}
	\caption{
		Radial profiles of the 4-speed $u(r)$ (left axis) and the magnetic field $B(r)$ (right axis) in the DF model (see Table \ref{BestFitParameter})
		for 3C 58 (left panel) and G21.5-0.9 (right panel).
		The dashed black line represents the radial profiles for the KC model.
		\vspace{5mm}
	}
	\label{fig:uB}
\end{figure*}

In Figure \ref{fig:uB}, the obtained radial profiles of 4-speed and the magnetic field are shown.
As expected from the value of $r_{\rm pe}/r_{\rm N}$,
$u(r)$ in 3C 58 shows an almost the same profile as the KC model profile.

In G21.5-0.9,
a modification of $u(r)$ due to the diffusion is seen.
For $r>10^{18}\cm$, since the magnetic pressure is large,
the deceleration and amplification of the magnetic field are almost saturated.
Despite the high-$\sigma$ and amplification of the magnetic field,
the magnitude of the average magnetic field is not so different between the DF model and the KC model,
because the volume-averaged magnetic field is largely controlled by the field near the edge of the nebula, which are almost the same
in the two models.

\highlight{
	For the DF model, obtained flow velocity at the edge of the nebula is $540$ km s$^{-1}$ and $720$ km s$^{-1}$ for 3C 58 and G21.5-0.9, respectively.
	These value are roughly consistent with the observed expansion speed $400$ km s$^{-1}$ and $870$ km s$^{-1}$ for 3C 58 \citep{2006ApJ...645.1180B} and G21.5-0.9 \citep{2008MNRAS.386.1411B}, respectively.
}

\begin{figure*}[!tb]
	\begin{center}
		$\begin{array}{cc}
		\includegraphics[width=0.5\textwidth]{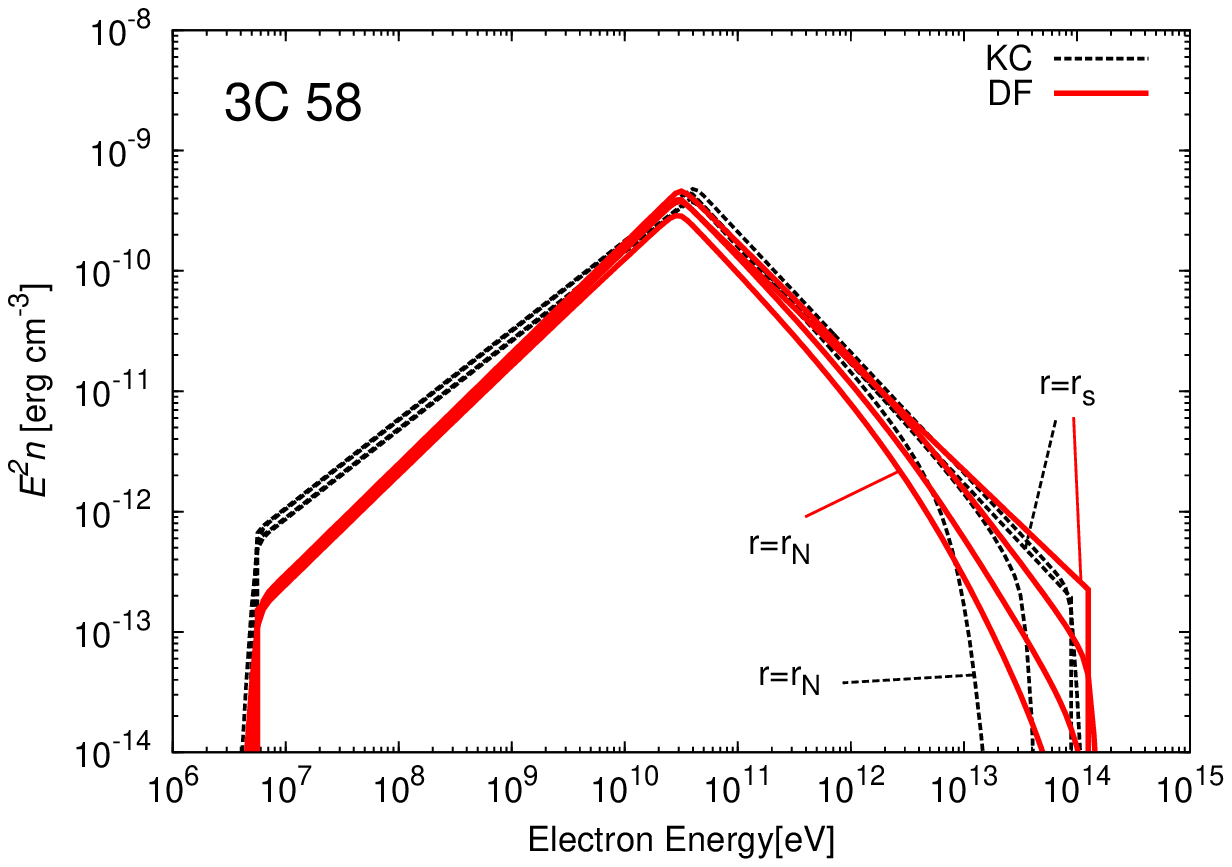} &
		\includegraphics[width=0.5\textwidth]{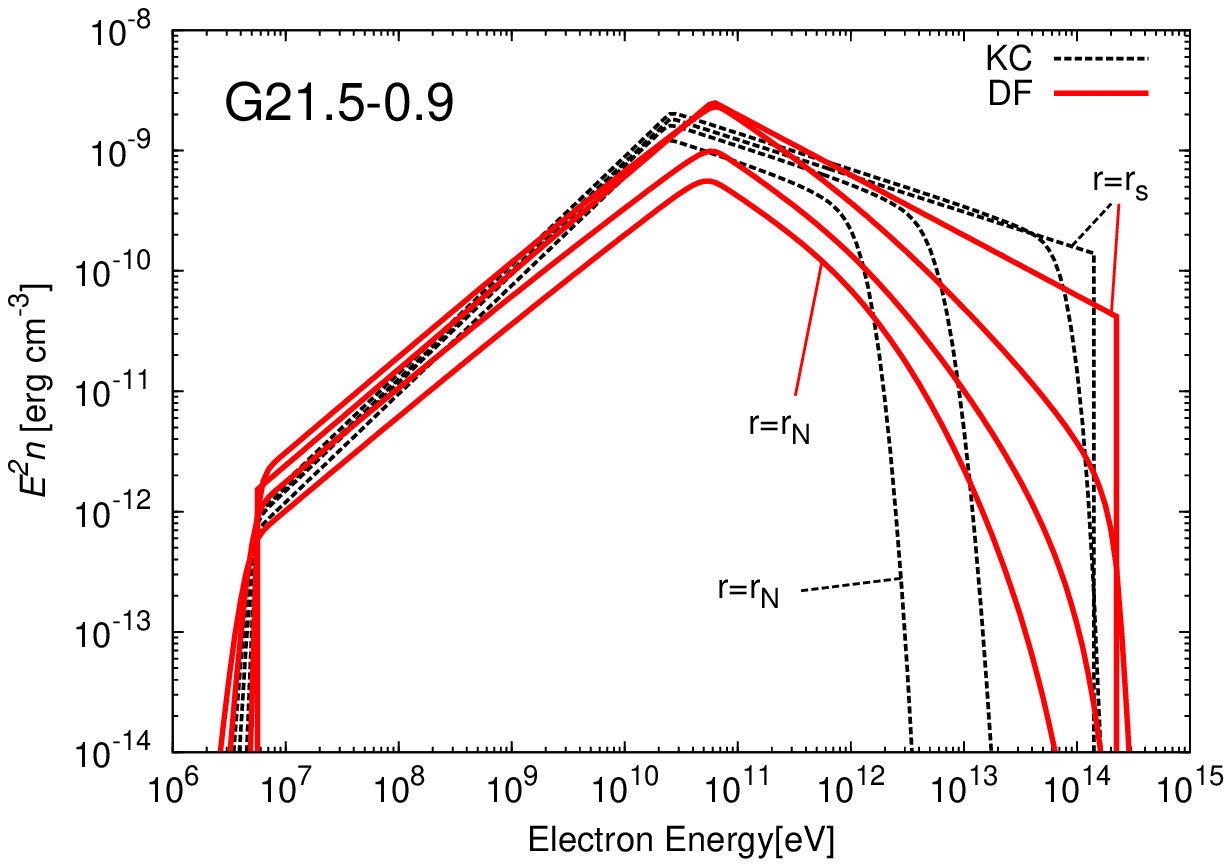}
		\end{array}$
	\end{center}
	\caption{
	The particle energy spectra for 3C 58 (left panel) and G21.5-0.9 (right panel) in the DF (red solid) and KC (black dashed) models.
	The different lines show the hard-to-soft evolutions: $r=r_{\rm s}$, $5 r_{\rm s}$, $10 r_{\rm s}$, $r_{\rm N}\simeq 15 r_{\rm s}$ for 3C 58,
	and $r=r_{\rm s}$, $6 r_{\rm s}$, $12 r_{\rm s}$, $r_{\rm N}= 18 r_{\rm s}$ for G21.5-0.9.
	\vspace{3mm}
	}
	\label{fig:electron}
\end{figure*}

\begin{figure*}[!tb]
	\begin{center}
		$\begin{array}{cc}
		\includegraphics[width=0.5\textwidth]{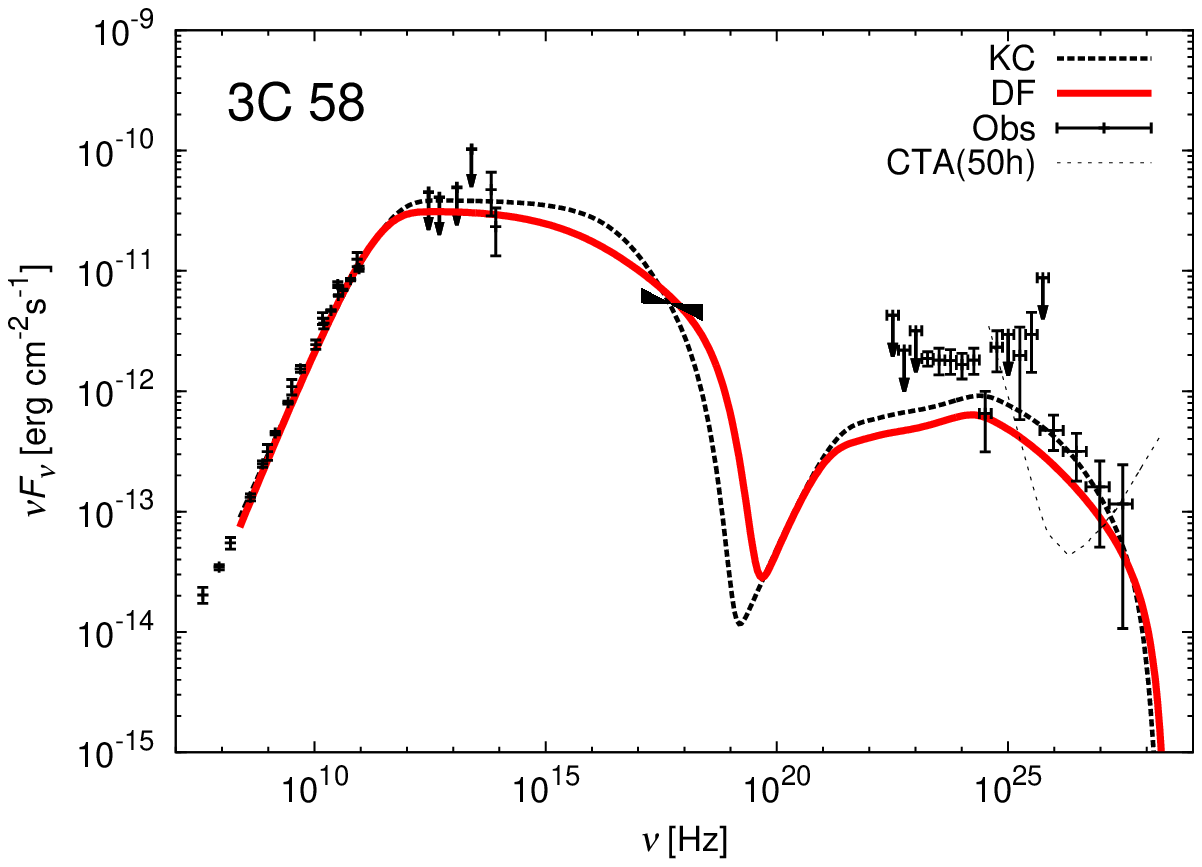} &
		\includegraphics[width=0.5\textwidth]{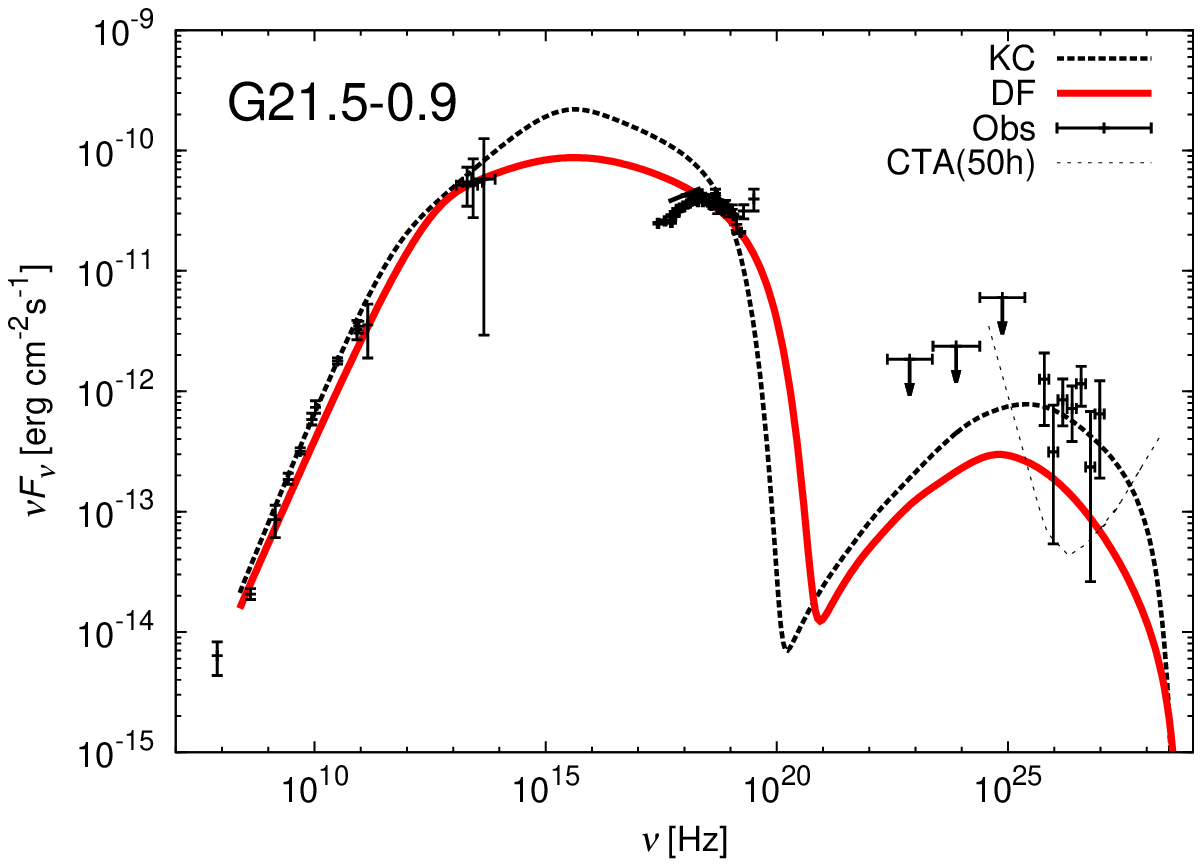}
		\end{array}$
	\end{center}
	\caption{
		Entire photon spectra for 3C 58 (left panel) and G21.5-0.9 (right panel).
		The red line represents the DF model, and the black line represents the KC model calculated by \citet{ishizaki17}.
		The data points are taken from \citet{2011ApJS..192...19W} (radio),
		\citet{1994ApJS...90..817G}, \citet{2008ApJ...676L..33S} (IR), \citet{2000PASJ...52..875T} (X),
		\citet{2018ApJ...858...84L} (GeV),
		and \citet{2014A&A...567L...8A} (TeV) for 3C 58,
		and \citet[][]{1989ApJ...338..171S} (Radio), \citet{1998MmSAI..69..963G} (IR),
		\citet{2011A&A...525A..25T}, \citet{2014ApJ...789...72N}, \citet{2009MNRAS.393..527D}, \citet{HitomiG21} (X),
		\citet{2011ApJ...726...35A} (GeV), and \citet{2008ICRC....2..823D} (TeV) for G21.5-0.9.
	}
	\label{fig:SED}
\end{figure*}

Figure \ref{fig:electron} shows the particle energy spectra for 3C 58 and G21.5-0.9,
and Figure \ref{fig:SED} shows the volume-integrated photon spectra.
First, for both the cases, the diffusion is not effective for particles at $ E'_\pm \sim E_{\rm min} $,
so the treatment with the fluid picture, as mentioned in Section \ref{sec:Model}, is justified.
In 3C 58,
while the diffusion is not effective for particles with  $ E'_\pm \sim E_{\rm b} $,
the particle spectrum above $10^{13} \eV$ is significantly modified by diffusion.
As discussed in Section \ref{sec:dependence},
the cutoff shape in the energy spectrum like the KC model does not appear in the DF model.
Since high energy particles propagate in the strong magnetic field region,
the DF model produces a harder entire spectrum.
As a result, the hard spectrum observed in X-ray is reproduced.
Thus, just considering the weak diffusion that does not affect the fluid motion, the entire spectrum changes greatly from the KC model.

In the DF model of G21.5-0.9, the diffusion is more effective than 3C 58.
As shown in Figure \ref{fig:uB}, the wind is decelerated more shallowly than $r^{-2}$ outside $r\sim 10^{18} \cm$, so the adiabatic cooling becomes efficient.
As a result, the energy spectrum of particles shows an evolution that all particles lose energy simultaneously with increasing radius.
Furthermore, the spectral peak of synchrotron radiation shifts to a higher frequency owing to the amplification of the magnetic field.
The highest frequency emission of the synchrotron component is originated from the radius where the magnetic field is most amplified,
while the IR/Optical emission is most strongly emitted near the edge of the nebula.
Particles near the edge of the nebula lose energy via adiabatic cooling, which
suppresses the IR/Optical flux in comparison with the X-ray flux.

In the DF model, the fluxes of $\gamma$-rays in both objects
are several times lower than the observation.
However, considering the following two points, this discrepancy can be solved.
The first is the uncertainty of the interstellar photon field.
The $\gamma$-rays are emitted mainly via the ICS with the infrared photon from the interstellar dust.
Since the flux of the ICS is roughly proportional to the energy density of the seed photons,
the $\gamma$-ray flux in our model is largely affected by the uncertainty of the inter stellar radiation field \citep[e.g.,][]{tor13}.
In addition, the photon field model used in our calculations is a value in ordinary interstellar space,
but the supernova remnants may be the site of dust formation \citep{2009ASPC..414...43K},
so that the intensity of the dust radiation may be different from ordinary interstellar space.

The second point is $\gamma$-ray emission from particles escaped out of the nebula.
Recently, \citet{2017ApJ...843...40A} (HAWC) reported that around the Geminga there is a much larger $\gamma$-ray halo than the X-ray PWN.
Even for MSH 15-52, whose age is comparable to that for G21.5-0.9 and 3C 58,
diffuse $\gamma$-ray emission extending beyond X-rays was detected \citep{2017arXiv170901422T}.
Those facts suggest that electrons and positrons escape out of the PWN to the ISM.
For example, an electron and a positron with energy $E_{\rm VHE}=1.5\times 10^{13} \eV$ product $\gamma$-rays of $\sim 1$ TeV via the ICS with the CMB photons ($\sim 10^{-3}$ eV).
For both fitting results, diffusion is more effective for such particles than advection.
If such particles continuously escape out of the nebula by diffusion\footnote{
	\highlight{
		While higher energy particles can escape from the nebula,
		almost all (low-energy) particles are still confined inside the nebula. 
	}
},
during the age of the nebula $\sim t_{\rm adv}$,
the number of escaped particles is $N_{\rm esc}\sim 4\pi r_{\rm N}^2 (\kappa(E_{\rm VHE})/r_{\rm N}) E_{\rm VHE} n(r_{\rm N},E_{\rm VHE}) t_{\rm adv}$.
Roughly adopting $E_{\rm VHE}^2 n=10^{-13}~\mbox{erg}~\mbox{cm}^{-3}$ ($10^{-12}~\mbox{erg}~\mbox{cm}^{-3}$) for 3C 58 (G21.5-0.9),
the number is estimated as $N_{\rm esc}\sim 10^{43}$ for both PWNe.
The energy release rate via ICS with the CMB photons is $\sim 10^{-11} \ {\rm erg}~{\rm s}^{-1}$ per particle.
Then, the $\gamma$-ray flux from escaped particles is estimated as comparable to the flux inside the nebula for both the cases.
The observed gamma-ray fluxes in Figure \ref{fig:SED} may include both the components inside and outside the nebula.
Assuming the same diffusion coefficient inside and outside the nebula (see section \ref{sec:Discussion}),
the extent of $\gamma$-rays is estimated as $\sim r_{\rm N}+\sqrt{\kappa t_{\rm adv}}$, which
is $\sim4 \pc$ ($400''$) for 3C 58, and $\sim 2 \pc$ ($90''$) for G21.5-0.9, respectively.

In G21.5-0.9, a spectral break around a few keV is reported by \citet{2014ApJ...789...72N} (NuSTAR) and \citet{HitomiG21}.
Furthermore, by using the time-dependent one-zone model,
\citet{HitomiG21} concluded that the spectral break is not reproduced under the conventional assumption of the particle injection and the energy loss processes.
If we try to explain this spectral break with our model,
we need to take parameters like the ``alternative model'' of \citet{ishizaki17},
in which the radio and IR/Opt emission are neglected and an extremely shorter advection time $t_{\rm adv}\sim40\ {\rm yr}$
than the pulsar age is required.
Since the cooling break becomes smooth due to the diffusion process in the DF model,
the sharp break in the X-ray region seems difficult to be reproduced.
If we seriously accept this spectral break, the alternative model without the diffusion in \citet{ishizaki17} would be better than the diffusion model.

\begin{figure*}[!tb]
	\begin{center}
		$\begin{array}{cc}
		\includegraphics[width=0.5\textwidth]{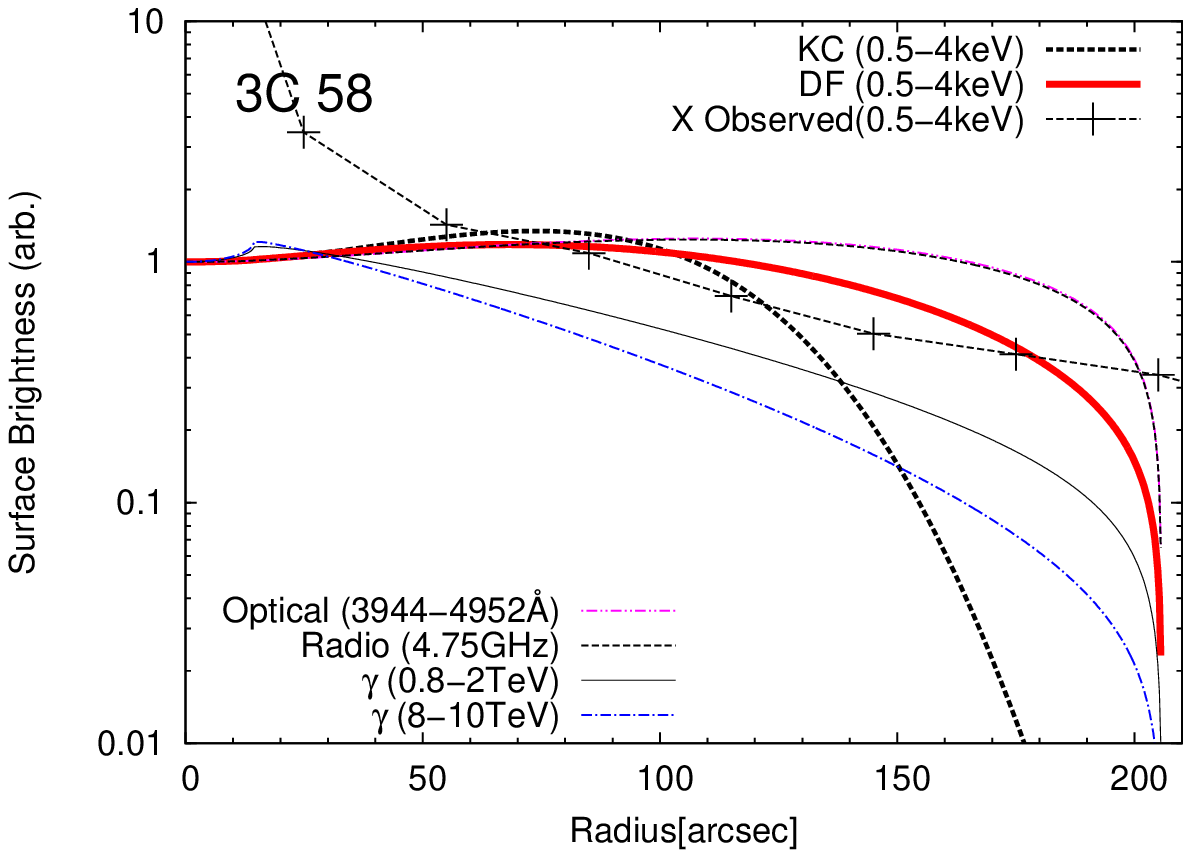} &
		\includegraphics[width=0.5\textwidth]{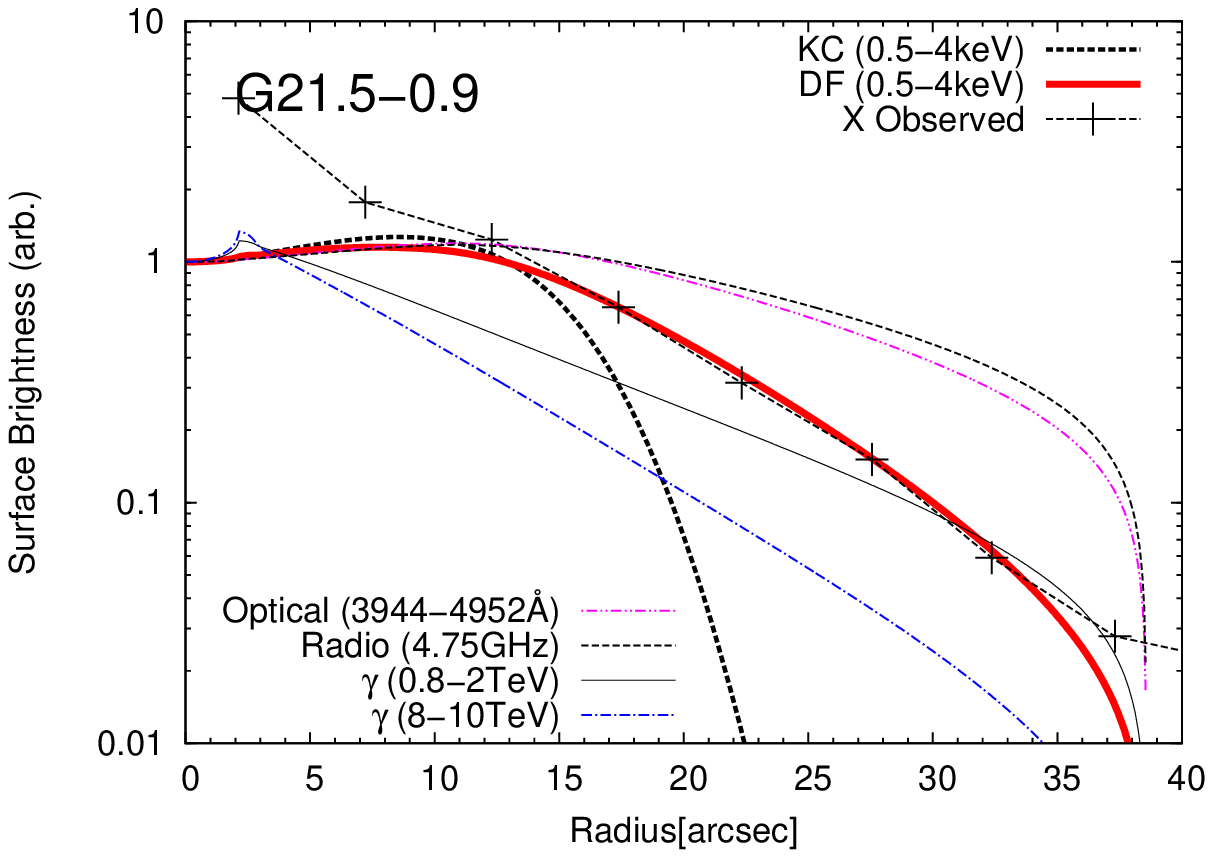}
		\end{array}$
	\end{center}
	\caption{
		Radial profiles of the surface brightness of X-rays (3-5 keV; bold), radio (4.75GHz; dashed),
		optical (3944-4952$\AA$; dashed double-dotted), 0.8-2 TeV $\gamma$-rays (thin) and 8-10 TeV $\gamma$-rays (dashed dotted).
		The red lines represent the DF model,
		and the thin black solid line represent the X-ray profile of the KC model.
		All curves are normalized as unity at the center.
		The data points are taken from \citet{2004ApJ...616..403S} (X) for 3C 58,
		and \citet{2001ApJ...561L.203B} (radio) and \citet{2005AdSpR..35.1099M} (X) for G21.5-0.9.
	}
	\label{fig:brightness}
\end{figure*}

\begin{figure*}[!tb]
	\begin{center}
		$\begin{array}{cc}
		\includegraphics[width=0.5\textwidth]{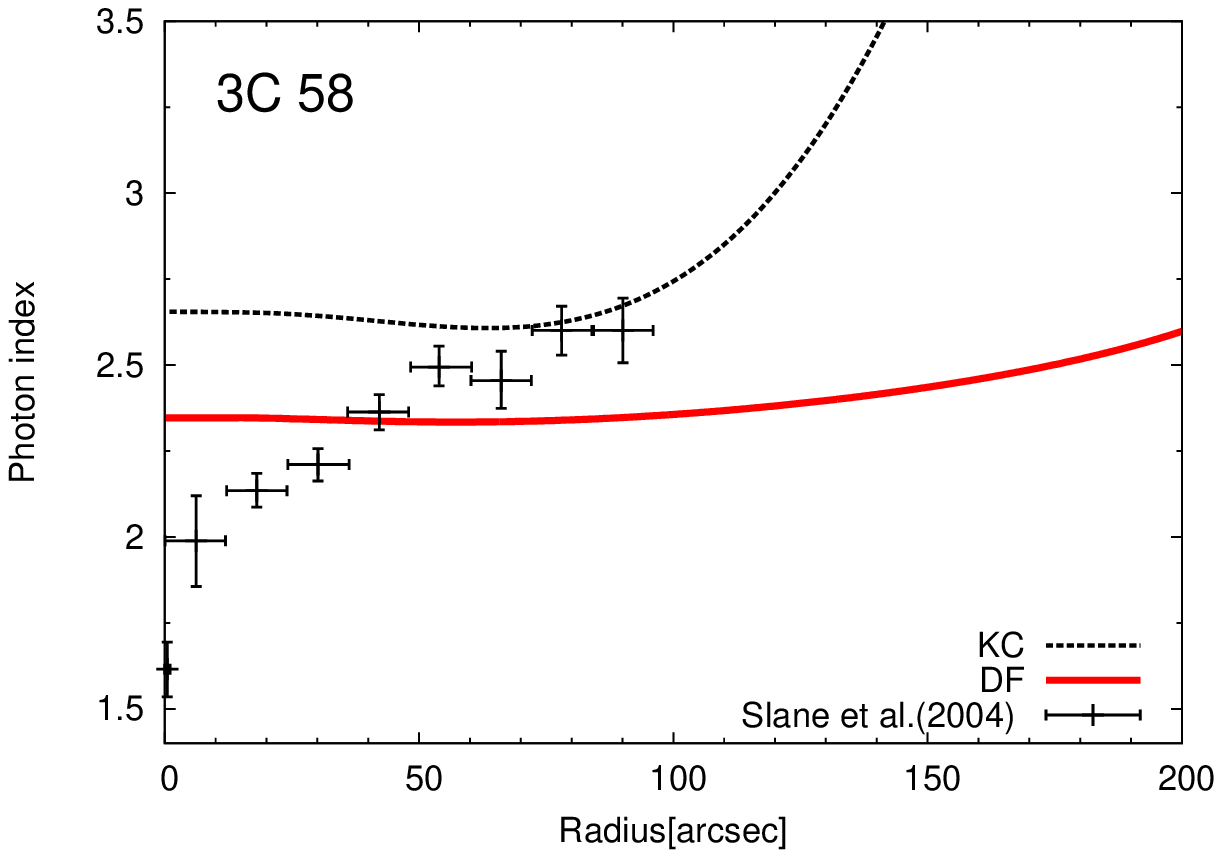} &
		\includegraphics[width=0.5\textwidth]{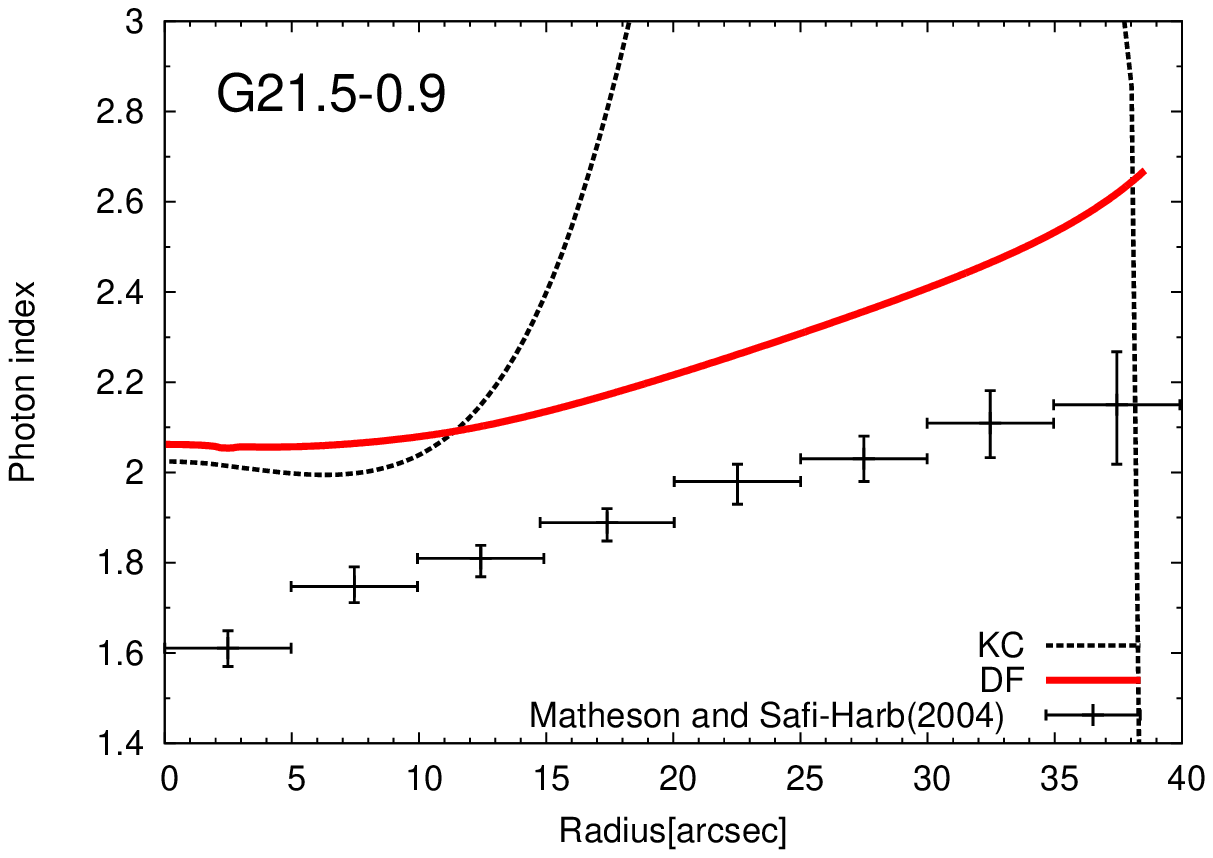}
		\end{array}$
	\end{center}
	\caption{
		Radial profiles of the photon indices in 0.5-10.0 keV range in the DF model (red), and the KC model (black).		
		The data points are taken from \citet{2004ApJ...616..403S}
		and \citet{2005AdSpR..35.1099M} for 3C 58 and G21.5-0.9, respectively.
		\vspace{6mm}
	}
	\label{fig:photon_index}
\end{figure*}

The radial profile of the X-ray surface brightness is shown with the red-thick line in Figure \ref{fig:brightness}.
The X-ray extent is large enough unlikely the KC model.
The observed X-ray profile of G21.5-0.9 is well reproduced quantitatively, but for 3C 58, the model curve is slightly deviated from the observation.
The X-ray brightness may be contaminated by emission from the SNR, and the spherically symmetric approximation may be not appropriate for both the observational data and the model.
We also plot the surface brightness in radio (4.75 GHz), optical (3944-4952 \AA), and $\gamma$-rays (0.8--2 TeV and 8--10 TeV).
For radio and optical band, the surface brightness profiles are almost spatially uniform with a sharp cut off at the edge.
For $\gamma$-rays, the width at half maximum becomes smaller with increasing energy of the observed $\gamma$-ray.
As we have discussed, if the particles escape from the nebulae by diffusion and emit $\gamma$-rays via ICS,
the extent of the $\gamma$-ray would be larger than the size of the nebula.

In Figure \ref{fig:photon_index}, the radial profile of the photon index is shown.
The problematic sudden softening in the KC model is not seen in the DF model.
Qualitatively the DF model greatly improves the index problem.
However, for 3C 58, the gradual softening in observed data seems inconsistent with almost the constant index in the DF model.
On the other hand, for G21.5-0.9, the gradual softening in the model is similar to the observation,
but the absolute value is slightly deviated.

\section{Summary and Discussion} \label{sec:Discussion}

In PWNe we cannot clearly distinguish the diffusive component from the background component.
Retaining the fluid picture,
we have developed a formulation to solve approximately the fluid dynamics with the spatial diffusion of particles self-consistently.
We have numerically calculated these equations for 1--D steady PWNe,
and investigated the dependence on the diffusion coefficient of the fluid structure (the velocity field and the magnetic field),
the entire spectrum, and the radial profile of the surface brightness in X-ray.
With increasing the diffusion coefficient, the entire spectrum becomes harder,
and the X-ray surface brightness extents larger.
For the fluid structure, we have found that the diffusion process causes the deceleration of the flow by the back reaction of diffusion.
The deceleration leads to amplification of the magnetic field, 
and influences the energy distribution of particles through radiation cooling nonlinearly.
When the diffusion coefficient is large,
the back reaction on the fluid dynamics significantly affects the emission spectrum.

We have applied this model to 3C 58 and G21.5-0.9.
This is the first attempt to reproduce both entire spectra and X-ray surface brightness with diffusion effect.
Roughly speaking, we have succeeded in reproducing them.
However, the $\gamma$-ray fluxes of entire spectra are the half darker than the observed ones,
and radial profiles of the photon index are slightly deviated from the observed data.

The escaped particles from the nebula also emit gamma-rays as detected from Geminga \citep{2017ApJ...843...40A},
where the diffusion coefficient outside the nebula is estimated as $\kappa(E=100\ {\rm TeV})\sim2\times10^{27} \cmcms$.
This value is close to the values inside the nebulae in our model.
Adopting the same diffusion coefficient outside the nebulae, our model
yields the gamma-ray halo of $400''$ for 3C 58, and $90''$ for G21.5-0.9, respectively.
The contribution from the halo emission may improve the dim gamma-ray fluxes in our model.
We expect that a better angular resolution of CTA \citep{2013APh....43..171B} will confirm an extended component
around G21.5-0.9 and 3C 58.

The diffusion coefficients at $E_\pm \sim10^{14}\eV$
in our model for G21.5-0.9 and 3C 58 are consistent with the previous values in \citet{por16} and \citet{tan12}.
In other words, the effect of the back-reaction of the diffusion does not affect the parameter estimate so much in those objects ($r_{\rm pe}/r_{\rm N} \gtrsim 1$).
However, it is not trivial whether the back-reaction affects significantly or not, in other cases.

The mean free path $\lambda$ at $E_\pm \sim10^{14}\eV$ in our model is obtained from $\kappa=\lambda c/3$ as $0.05 \pc$ in both the nebulae.
Since we have assumed a toroidal magnetic field, the diffusion of particles is perpendicular to the global magnetic field in our model.
In this case, the mean free path is given by $\lambda\sim r_{\rm L}(\delta B/B)^2$ in the framework of the quasi-linear theory of the resonant scattering
\citep[e.g.,][]{1987PhR...154....1B},
where $r_{\rm L}$ is the gyro radius of the particle, and $\delta B$ is the turbulence amplitude at the scale of $\sim r_{\rm L}$.
\highlight{
	Our model parameter set implies $r_{\rm L}\sim 10^{-3} \pc$ so that $\delta B/B > 1$ is required to achieve the value $\kappa\sim 10^{27}\cmcms$ at $10^{14}$ eV,
	which indicates that the simple perpendicular diffusion theory is not applicable
}
(cf. in the typical interstellar space, $\delta B/B\sim O(10^{-2})$ \footnote{This is estimated assuming the parallel diffusion \citep[e.g.,][]{2007ARNPS..57..285S}.}).
As proposed in \citet{tan12},
if the direction of the magnetic field is deformed in the radial direction as seen in the Rayleigh-Taylor finger, 
the particles may be able to diffuse efficiently toward the radial direction.
On the other hand, the transport of particles due to the global eddy motion in the nebula,
as \citet{por16} discussed, is also a candidate for the origin of the efficient diffusion.
In this case, the energy dependence of the diffusion coefficient becomes much weak,
and such a model is different type from ours.
To treat the particle escape from the nebula, another assumption other than the global eddy motion is additionally required in this model.
Thus, the value of the diffusion coefficient our model suggests is not trivial.
There may be another option to modify the KC model other than diffusion.

As shown in Figure \ref{fig:photon_index}, 
the photon spectral index in the DF model of G21.5-0.9 is larger than the observed value,
but the gradual softening in the radial direction is consistent with the observation.
As we discussed in Section \ref{sec:application},
it is difficult to reproduce the observed spectral break of several keV with the effects of cooling and diffusion.
Those facts suggests that there may be multiple components in the particle spectrum.
Radio-emitting particles and X-ray emitting particles may have different origins,
as proposed by \citet{ishizaki17} and \citet{2017ApJ...841...78T}.
An X-ray and gamma-ray observation focusing on the morphology coincidence in radio and IR/optical will provide a clue to revealing those origins.

In this model, for simplicity, we have adopted a uniform diffusion coefficient.
If we consider the spatial dependence of the diffusion coefficient,
a more sophisticated model to explain the observed properties may be possible.
In order to establish such a model, study for both the microscopic plasma waves
and hydrodynamic turbulence should be promoted.
Moreover, a detailed investigation of the escape process from the nebula is also required.
Since particles escape through the contact discontinuity to the SNR,
the turbulence near $r=r_{\rm N}$ as the result of the interaction between the SNR and the PWN is important.
In steady outflow models,
the hydrodynamical structure at the edge of the nebula cannot be examined.
The vicinity of the edge of the nebula has a large volume,
so that this kind of uncertainty may greatly affect the emission spectrum.
A time-dependent 1--D model will clarify the contribution of emission from near the edge of the nebula.

\section*{Acknowledgments}
We first appreciate the referee for valuable comments.
We are grateful to Shuta Tanaka for useful discussion.
This work is supported by Grants-in-Aid for Scientific
Research No. 16K05291 (KA) from the Ministry
of Education, Culture, Sports, Science and Technology (MEXT) of Japan.
This work is carried out by the joint research program of the Institute for Cosmic Ray Research (ICRR), The University of Tokyo.


\bibliography{draft}

\end{document}